\documentclass[journal]{IEEEtran}
\usepackage{xcolor,soul,framed} 
\colorlet{shadecolor}{yellow}
\usepackage[pdftex]{graphicx}
\graphicspath{{../pdf/}{../jpeg/}}
\DeclareGraphicsExtensions{.pdf,.jpeg,.png}
\usepackage[cmex10]{amsmath}
\usepackage{array}
\usepackage{amssymb}
\usepackage{mdwmath}

\usepackage{mdwtab}
\usepackage{eqparbox}
\usepackage{url}
\usepackage{enumerate}
\usepackage{enumitem}
\usepackage{amsfonts}
\usepackage[english]{babel}
\usepackage[utf8]{inputenc}
\usepackage[noend]{algpseudocode}
\usepackage{mathtools}
\usepackage{nomencl}
\usepackage{blindtext}
\usepackage{amsmath}
\usepackage[colorinlistoftodos]{todonotes}
\usepackage{color}
\usepackage{color,soul}
\usepackage{xcolor}
\usepackage{multirow}
\usepackage{graphicx}
\usepackage{epstopdf}
\usepackage[T1]{fontenc}
\usepackage{textcase}
\usepackage{array}
\usepackage{mdwmath}
\usepackage{mdwtab}
\usepackage{eqparbox}
\usepackage{enumerate}
\usepackage{bm}
\usepackage{amsfonts}
\usepackage[ruled,lined,linesnumbered]{algorithm2e}
\usepackage{booktabs}
\usepackage{comment}
\usepackage{amsmath}
\usepackage{mathtools}
\usepackage{caption}
\usepackage{subcaption}
\usepackage{tabularx}
\usepackage{hyperref}
\hypersetup{
    colorlinks=true,
    linkcolor=blue,
    filecolor=magenta,      
    urlcolor=cyan,
}
\usepackage{cleveref}

\usepackage{subcaption}
\usepackage{caption}
\usepackage{mathrsfs}
\usepackage{booktabs}
\usepackage{siunitx}
\usepackage{pythonhighlight}

\usepackage{cite}
\usepackage{url}
\usepackage{setspace}
\usepackage{amsthm}
\usepackage{nomencl}
\usepackage{footnote}
\usepackage{threeparttable}
\usepackage{tikz}
\usepackage{tabularx}

\usepackage{wrapfig}

\makenomenclature
\usepackage{etoolbox}
\renewcommand\nomgroup[1]{%
  \item[\bfseries
  \ifstrequal{#1}{A}{Abbreviations}{%
  \ifstrequal{#1}{V}{Vectors, Matrices, and Sets}{%
  \ifstrequal{#1}{O}{Operators}{%
  \ifstrequal{#1}{S}{Operators, Sets, and Symbols}{}}}}%
]}
\usepackage{pifont}
\usepackage[normalem]{ulem}
\newcommand{\cmark}{\ding{51}} 
\newcommand{\xmark}{\ding{55}} 

\newtheorem{definition}{Definition}
\newtheorem{remark}{Remark}

\tikzset{square arrow/.style={to path={-- ++(0,-.25) -| (\tikztotarget)}}}

\usepackage{xcolor,cite,etoolbox}
\usepackage{tabularx}
\makeatletter 
\pretocmd\@bibitem{\color{black}\csname keycolor#1\endcsname}{}{\fail}

\usepackage{etoolbox}
\makeatletter
\patchcmd{\@makecaption}
  {\scshape}
  {}
  {}
  {}
\makeatother

\title{Learning-Augmented Power System Operations: \\ A Unified Optimization View} 

\author{Wangkun Xu,~\IEEEmembership{Member,~IEEE}, Zhongda      
    Chu,~\IEEEmembership{Member,~IEEE}
    and Fei Teng,~\IEEEmembership{Senior Member,~IEEE}\\
    \thanks{
    Wangkun Xu and Fei Teng (\{wangkun.xu18,f.teng\}@imperial.ac.uk) are with the Department of Electrical and Electronic Engineering, Imperial College London, UK; Zhongda Chu (z\_chu@tju.edu.cn) is with Tianjin University, China. This work was supported by the Engineering and Physical Sciences Research Council grant number [EP/Y025946/1]. (\emph{Corresponding author: Fei Teng})
    }
}
\begin{document}

\renewcommand{\baselinestretch}{1}
\markboth{Submitted to IEEE}%
{Shell \MakeLowercase{\emph{et al.}}: Bare Demo of IEEEtran.cls for IEEE Journals}
\maketitle

\begin{abstract}
With the increasing penetration of renewable energy and inverter-based resources, traditional physics-based power-system operation faces growing challenges in maintaining economic efficiency, security, and robustness. Machine learning (ML) has emerged as a powerful tool for modeling complex system dynamics and uncertainty. However, standalone ML pipelines, including model selection, training, and validation, are often designed separately from the downstream optimization problems they influence, which can lead to suboptimal system-level decisions. To address this gap, this paper proposes \emph{Learning-Augmented Power System Operations} (LAPSO), a unified optimization-centered framework that treats ML as an explicit component of power-system operational decision-making. First, LAPSO provides generalized mathematical template covering both decision-independent predictors that parameterize downstream optimization and decision-dependent learned surrogates that enter optimization as auxiliary constraints. Second, it designs ML pipelines using optimization-aware criteria, including solution-quality, computational tractability, constraint satisfaction, and economic performance. We instantiate LAPSO on both stability-constrained optimization (SCO) and objective-based forecasting (OBF), and show how the framework provides actionable guidance for selecting learned components. We further extend the framework to a hybrid forecast--operation--control chain and use it to organize heterogeneous uncertainty sources. Finally, we release an open-source Python package, \texttt{lapso}, for modularly augmenting existing power-system optimization models with ML components. Code and datasets are available at: \url{https://github.com/xuwkk/lapso_exp}.
\end{abstract}


\setlength{\nomlabelwidth}{1.5cm}   

\nomenclature[A]{TPR/FPR}{True/False Positive Rate}
\nomenclature[A]{IBR}{Inverter-based Resources}
\nomenclature[A]{UC/ED}{Unit Commitment/Economic Dispatch}
\nomenclature[A]{IBP}{Interval Bound Propagation}
\nomenclature[A]{DP/RD}{Dispatch/Redispatch}
\nomenclature[A]{OPF}{Optimal Power Flow}
\nomenclature[A]{NN}{Neural Network}
\nomenclature[A]{SCO}{Stability-constrained Optimization}
\nomenclature[A]{OBF}{Objective-based Forecasting}
\nomenclature[A]{LAPSO}{Learning-augmented Power System Operations}
\nomenclature[A]{ABF}{Accuracy-based Forecasting}
\nomenclature[A]{ID}{Implicit Differentiation}
\nomenclature[A]{gSCR}{Generalized Short Circuit Ratio}
\nomenclature[A]{(c)LgR}{(Constrained) Logistic Regression}
\nomenclature[A]{LR}{Linear Regression}
\nomenclature[A]{BCE}{Binary Cross-Entropy}
\nomenclature[A]{MIL(P)}{Mixed-Integer Linear (Program)}
\nomenclature[A]{ML}{Machine Learning}
\nomenclature[A]{MP(EC)}{Mathematical Program (with Equilibrium Constraints)}
\nomenclature[A]{KKT}{Karush-Kuhn-Tucker}
\nomenclature[A]{SGD}{Stochastic Gradient Descent}
\nomenclature[A]{AD}{Automatic Differentiation}
\nomenclature[A]{C\&CG}{Column and Constraint Generation}

\nomenclature[S]{$\operatorname{diag}(\cdot)$}{Diagonalization on a vector}
\nomenclature[S]{$(\cdot)^T$}{Matrix transpose}
\nomenclature[S]{$\mathbb{P}(\cdot)$}{Probability distribution}
\nomenclature[S]{$\partial_\divideontimes(\cdot)$}{Partial differentiation with respect to $\divideontimes$ element}
\nomenclature[S]{$\mathbb{E}(\cdot)$}{Expectation operator}
\nomenclature[S]{$\mathcal{KKT}(\cdot)$}{Set of KKT conditions}
\nomenclature[S]{$\circ$}{Hadamard product}

\nomenclature[S]{$\bm{z}$}{Decision variables of an optimization program}
\nomenclature[S]{$\bm{y}_1,\bm{y}_2$}{Exogenous and predicted parameters}
\nomenclature[S]{$\widehat{(\cdot)}$}{Forecasted variable}
\nomenclature[S]{$(\cdot)^\star$}{Optimal variables}
\nomenclature[S]{$\bm{\theta}$}{ML parameter}
\nomenclature[S]{$\bm{x}$}{Contextual input of ML model}
\nomenclature[S]{$\widetilde{(\cdot)}$}{Modified variable or function}
\nomenclature[S]{$\bm{v}(\cdot;\bm{\theta})$}{ML model for abstract LAPSO problem}
\nomenclature[S]{$u(\cdot;\bm{\theta})$}{ML model for SCO problem}
\nomenclature[S]{$\bm{h}(\cdot;\bm{\theta})$}{ML model for OBF problem}
\nomenclature[S]{$f(\cdot)$}{Objective function of an optimization}
\nomenclature[S]{$\bm{g}(\cdot)$}{Constraint functions of an optimization}
\nomenclature[S]{$\mathcal{D}$}{Training dataset}
\nomenclature[S]{$\mathcal{C}(\cdot)$}{Set of constraints}
\nomenclature[S]{$\textbf{NN}_{\bm{L}}^{\bm{\psi}}$}{NN with $\bm{L}$ number of linear layers and $\bm{\psi}$ number of trainable parameters.}
\nomenclature[S]{$\mathcal{P}$}{Power system optimization.}
\nomenclature[S]{$\mathcal{M}$}{ML training optimization.}

\printnomenclature

\section{Introduction}

\subsection{Background and Motivation}

\begin{figure}
    \centering
    \includegraphics[width=0.95\linewidth, trim={0.0cm 0.0cm 0cm 0}, clip]{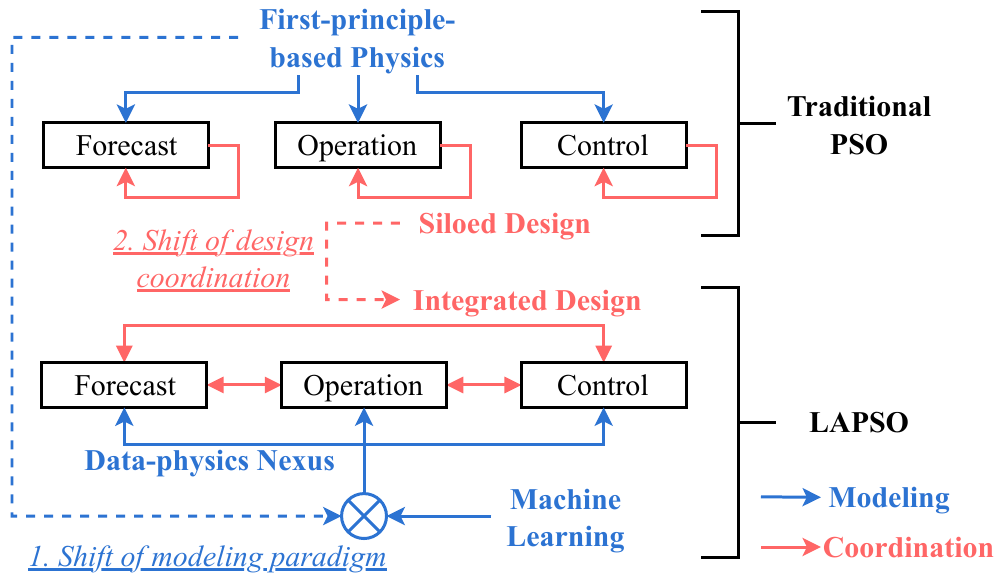}
    \caption{Transformation of power system operational methods.}
    \label{fig:decision_making}
\end{figure}

Power system operational decision-making consists of sequentially connected tasks, such as forecasting, operation, and control (Fig.~\ref{fig:decision_making}) \cite{conejo2018power}. With the decarbonization need, traditional physics-based approaches face growing challenges in maintaining economic efficiency, stability, and robustness. For example, increasing renewable uncertainty undermines deterministic operations based on point forecasts \cite{aien2016comprehensive}. Meanwhile, the proliferation of inverter-based resources (IBRs) can introduce small-signal stability issues through phase-locked-loop interactions, and reduced system inertia limits frequency support following disturbances \cite{cui2025control}. As a result, system operator must explicitly manage uncertainty and avoid instability, yet the high-order dynamics of both renewable generation and grid dynamics are hard to describe in analytical form or to assess efficiently in real time. Machine/deep learning (ML/DL) has therefore emerged as a promising option for renewable forecasting \cite{wang2019review} and stability assessment \cite{zhang2023data}, due to its representational capacity and efficient online inference \cite{bellizio2022transition}. These applications treat \emph{ML as a modeling tool} for complex, partially observed system components, reflecting \emph{a shift toward a data--physics nexus} in operational pipelines (Fig.~\ref{fig:decision_making}, red).

Despite this progress, conventional decision-making remains largely \emph{siloed}: forecasting, operational optimization, and control are often designed independently, even though their decisions are causally coupled. From an operational perspective, accuracy-driven forecasting overlooks its cascading impact on downstream operational cost and feasibility \cite{zhang2025decision}. Conversely, cost-driven operational decision ignores stability requirements that manifest at the control stage, leading to either insufficient security margins or overly conservative operation \cite{cui2025control, xu2024incorporation}. From a learning perspective, developing ML models in isolation neglects that the ML output becomes a learnable component of the power system operation (PSO) pipeline. Therefore, model structure, training objectives, and validation criteria should reflect not only predictive accuracy but also the system-level outcomes induced by the deployed optimization. This motivates a second transformation: \emph{shifting from siloed pipeline design toward an integrated learning--optimization framework} that supports end-to-end optimization-aware evaluation (Fig.~\ref{fig:decision_making}, blue).

To clarify the idea, we focus on two canonical integration patterns between data-driven learning and physics-based optimization, namely stability-constrained optimization (SCO) and objective-baesd forecasting (OBF). These two coupling modes induce both shared and distinctive implications for integrated learning-optimization design, yet are usually studied separately.


\subsection{Literature Review}

\subsubsection{SCO} 

SCO incorporates data-driven stability assessments into cost-driven power-system optimization to enforce adequate security margins \cite{zhang2023data}. A common motivation is to replace expensive dynamic simulation and stability analysis (e.g., ODE-based assessments) with an algebraic surrogate that can be embedded as constraint in real-time optimizations. Existing studies have explored a range of stability criteria and model classes, including conic regression for small-signal stability-constrained unit commitment (UC) \cite{chu2023stability}; input-convex neural networks and feedforward neural networks for transient or frequency stability-constrained UC \cite{wu2023transient, xia2024efficient, wang2025regional, tuo2022deep}; decision trees for frequency-stability constrained AC OPF \cite{chen2024adaptive}; and learning multiple convex voltage-stability constraints \cite{jia2025learning}. 

\subsubsection{OBF} 

OBF aims to align forecasting with downstream decision-making objectives by treating the optimization problem as an implicit loss function within the ML training loop \cite{zhang2025decision}. In exact formulations, the downstream optimization is solved and sensitivity information on operational cost is used to update forecaster parameters \cite{munoz2022bilevel, chen2024towards}. For neural forecasters trained with stochastic gradient descent (SGD), gradient-based implementations typically rely on differentiable optimization or implicit differentiation (ID) \cite{blondel2024elements} to compute sensitivities of the optimal solution with respect to forecaster parameters, as demonstrated in OBF applications for economic dispatch and related settings \cite{vohra2023end, xu2024e2e, wahdany2023more}. To reduce training-time computational burden, surrogate approaches have been explored, by quantifying the costs induced by forecast error using piecewise-linear surrogates \cite{zhang2022cost}, or approximating value functions of nonconvex dispatch--redispatch problem using neural networks embedded within the forecast model for fine-tuning \cite{beichter2024decision}. 

\subsubsection{Uncertainty in SCO and OBF} 

As ML models (especially neural networks) are often treated as black-boxes, uncertainties in SCO and OBF have been actively discussed. From the robustness perspective, \cite{chen2023vulnerability} analyzes the worst-case impact of uncertainty in the inertia forecaster on operational cost, while \cite{xu2024e2e} employs an adversarial training approach to improve robustness. In stochastic settings, \cite{donti2017task} incorporates predictive uncertainty by formulating OBF training with a stochastic dispatch problem as the downstream optimization. For SCO, \cite{zuo2024transferability} studies stability misclassification induced by uncertainty in input features at the operation stage, and \cite{chu2024managing} further explores the impact of uncertain grid parameters on data-driven stability constraints. 

\subsection{Contributions}
Despite the growing adoption of ML in power system operations such as SCO and OBF, existing learning--optimization methods are often developed case-by-case, and the relationship between ML design choices and downstream operational performance, including the robustness assessment under uncertainty, remains opaque without a unified framework. 
Broadly speaking, we ask:
\begin{quote}
\emph{Can we establish a general framework for designing learning models that integrate with existing power system optimizations, with explicit guidance on model selection, training, and validation under system-level objectives and computational constraints?}
\end{quote}
This paper addresses this question by proposing a unified design framework, with the following contributions.

\begin{enumerate}[leftmargin=*]
    \item \textbf{Unified framework.}
    We propose \emph{Learning-Augmented Power System Operations} (LAPSO), a unified framework that treats ML models as explicit components of optimization-centered power-system operations, and provides mathematical template for both decision dependent and independent coupling. 

    \item \textbf{Optimization-aware metrics.}
    Beyond the predictive metrics, we introduce an optimization-aware design view for ML in power systems, organized around solution-quality indicators, computational tractability, constraint satisfaction, and economic performance, providing explicit guidance for model selection, training, and validation once learning is embedded into operational optimization.

    \item \textbf{Demonstration on representative integration.}
    We first instantiate LAPSO on two existing representative applications: SCO and OBF, showing how different learning--optimization couplings follow the same LAPSO principle but induce different design preferences. We further extend the applications to a hybrid forecast--operation--control setting to optimize end-to-end system performance.


    \item \textbf{Uncertainty quantification.} Using LAPSO pipeline, we distinguish uncertainty sources in predicted quantities from exogenous optimization parameters, and discuss how uncertainties propagate to optimization-aware outcomes. We further provide sensitivity-based diagnostics to assess transferability of ML model across different optimization settings. Moreover, a novel multi-stage robust ML training algorithm is proposed to hedge against wait-and-see uncertainties on exogenous optimization parameters.

    \item \textbf{Open-source implementation.}
    We release open-source Python packages: \texttt{gridforge} for scalable generation of power-system testbeds and associated grid-aware datasets, and \texttt{lapso} for modularly augmenting existing power-system optimization models with learned components and solver-compatible encodings. The effectiveness of LAPSO and the accompanying packages are evaluated on IEEE 14-, 39-, 57-, 118-, and 300-bus systems for various tasks.
\end{enumerate}

This paper is organized as follows.
The proposed LAPSO pipeline and design metrics are introduced in Section~\ref{sec:lapso_framework}. Its applications to SCO and OBF are discussed in Section~\ref{sec:special_type}. New closed-loop forecast-operation-control applications and uncertainty quantification are presented in Section~\ref{sec:new_application}. An introduction to the \texttt{lapso} package is provided in Section~\ref{sec:package}. The case study settings, including detailed mathematical formulations, new robust algorithms, and the implementation of \texttt{lapso} package, are described in Section~\ref{sec:case_study}. The simulation results are presented in Section~\ref{sec:simulation}. Section~\ref{sec:conclusion} concludes the paper with the scalability analysis in the Appendix.

\section{The LAPSO Framework}\label{sec:lapso_framework}

\subsection{Formulation}\label{sec:lapso_formulation}

\begin{figure}[t]
    \centering
    \includegraphics[width=0.8\linewidth, trim={0.0cm 0cm 0cm 0}, clip]{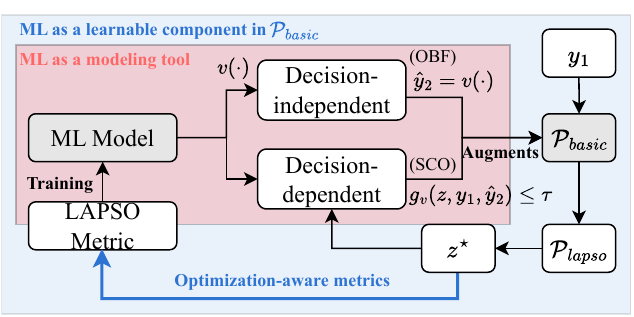}
    \caption{Visualization on LAPSO framework.}
    \label{fig:lapso_scope}
\end{figure}

Power-system operational decision-making consists of sequentially connected tasks, e.g. forecasting, operation, and control (Fig.~\ref{fig:decision_making}). This paper considers a class of optimization problems, each of which can be compactly written as
\begin{equation*}
        \boxed{\mathcal{P}_{basic}}:\quad \min_{\bm{z}} \; f(\bm{z};\bm{y}) \quad
        \text{s.t.} \; \bm{g}(\bm{z};\bm{y}) \leq \bm{0}.
\end{equation*}
where $\mathcal{P}_{basic}$ is considered as \emph{basic} operational problem currently implemented in practice; $\bm{z}$ denotes the operational decision variables and $\bm{y}$ denotes the problem parameters, such as load and renewable profiles or market prices. In developing the LAPSO framework, we do \emph{not} restrict $\mathcal{P}_{basic}$ to a specific optimization class. For example, $\mathcal{P}_{basic}$ may represent day-ahead unit commitment (UC) or dispatch (DP), near-real-time economic dispatch (ED), redispatch (RD), or optimal power flow (OPF). 

To ensure grid stability and robust operation, ML-based surrogate models are introduced as the new component to model the uncertainties in renewable productions and complex IBR dynamics. Such conventional learning framework is referred as \emph{standalone ML design} and to better align the ML model with $\mathcal{P}_{basic}$ at both the training and deployment stages, this paper proposes the concept of \emph{learning-augmented operation}.
\begin{definition}
    A power system operation $\mathcal{P}_{basic}$ is learning-augmented if some of its components are represented by parametric models $\bm{v}(\cdot;\bm{\theta}^\star)$.
\end{definition}
Mathematically, it extends $\mathcal{P}_{basic}$ to:
\begin{equation*}
    \begin{aligned}
        \boxed{\mathcal{P}_{lapso}}: \min_{\bm{z}}\; & f(\bm{z};\bm{y}_1,\hat{\bm{y}}_2) + f_v(\bm{z};\bm{y}_1,\hat{\bm{y}}_2) \\
        \text{s.t.}\; & \bm{g}(\bm{z};\bm{y}_1,\hat{\bm{y}}_2) \leq \bm{0} \\
        & \bm{g}_v(\bm{z};\bm{y}_1,\hat{\bm{y}}_2)\leq \bm{\tau} \\
        & \hat{\bm{y}}_2 = \bm{v}(\bm{z},\bm{y}_1,\bm{x};\bm{\theta}^\star).
     \end{aligned}
\end{equation*}
As shown by Fig.~\ref{fig:lapso_scope}, the parameter $\bm{y}$ is classified into \emph{exogenous} and \emph{predicted} parameters as $\bm{y}_1$ and $\bm{y}_2$, respectively. $\bm{y}_1$ is treated as given inputs to $\mathcal{P}_{lapso}$, which is not modeled by ML predictor and $\bm{y}_2$ is allowed to contain auxiliary variables that are not included in the original $\bm{y}$.
In the supervised setting, a parametric model $\bm{v}(\cdot;\bm{\theta})$ is trained over the dataset $\{(\bm{x},\bm{z},\bm{y}_1), \bm{y}_2\} \in \mathcal{D}$ under a suitable loss function. The feature or model input includes $\bm{y}_1$, $\bm{z}$, and contextual information $\bm{x}$. Apparently, $\bm{v}(\cdot;\bm{\theta})$ can represent either regression or classification models. After training, the selected model parameter is denoted as $\bm{\theta}^\star$, and the model forecasted parameter is denoted as $\hat{\bm{y}}_2$.

The terms $f_v(\cdot)$ and $\bm{g}_v(\cdot)\leq\tau$ are introduced as \emph{generic placeholders} for how learned components enter $\mathcal{P}_{basic}$. Depending on the operational requirement, ML modules can be coupled to the optimization layer in two canonical ways: (i) \emph{soft coupling} through an objective contribution $f_v(\cdot)$, and (ii) \emph{hard coupling} through explicit constraints $\bm{g}_v(\cdot)\le\tau$. In detail, safety or stability-critical specifications are naturally modeled as hard or risk-averse constraints, whereas cost trade-offs, e.g. preferences and regularization, are modeled as penalties in the objective.

Moreover, LAPSO is defined to encompass both \emph{decision-independent} predictive models, where $\bm{v}(\cdot)$ does not take decision variables $\bm{z}$ as input, and \emph{decision-dependent} predictive models, where the learned component explicitly depends on decisions. The latter arises in hybrid learning--optimization settings, e.g., when learning decision-dependent system responses, surrogate constraints, or post-decision dynamics.

A more compact form with $\widetilde{(\cdot)}$ representing the modified version of $f$ or $\bm{g}$ in $\mathcal{P}_{basic}$ can be denoted as
\begin{equation*}
    \boxed{\mathcal{P}_{lapso}}: \min_{\bm{z}} \; \tilde{f}(\bm{z};\bm{y}_1,\bm{x},\bm{\theta}^\star) \quad 
    \text{s.t.} \; \tilde{\bm{g}}(\bm{z};\bm{y}_1,\bm{x},\bm{\theta}^\star) \leq \bm{0},
\end{equation*}
where the objective and constraints are extended to include machine learning parameters $\bm{\theta}^\star$.

\subsection{Standalone ML Pipeline}\label{sec:lapso_framework_ml}

\begin{figure}[t]
    \centering
    \includegraphics[width=0.8\linewidth]{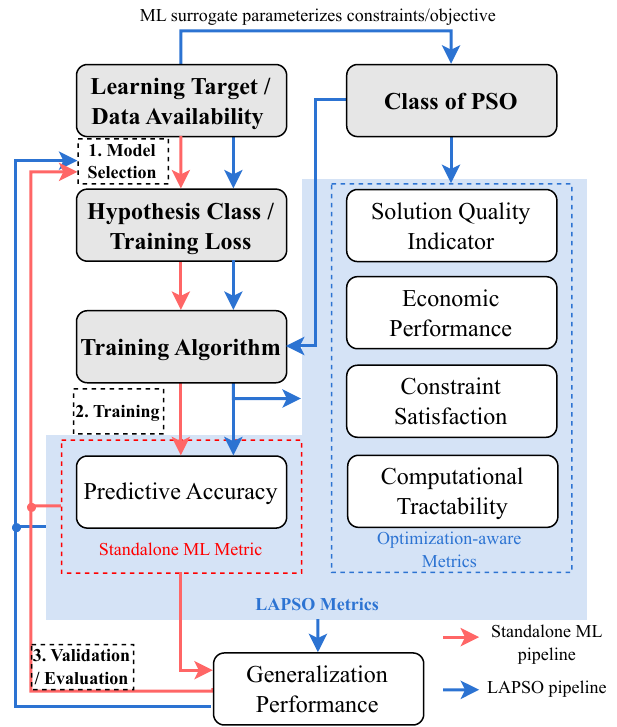}
    \caption{Training pipelines of standalone ML and LAPSO under predictive accuracy and optimization-aware system metrics.}
    \label{fig:learning_pipeline}
\end{figure}


As shown in Fig.~\ref{fig:learning_pipeline}, the ML workflow consists of \emph{model selection}, \emph{training}, and \emph{validation} with respect to a specified learning target and evaluation criteria. We summarize the design primitives in ML \cite{tuo2022deep} into: (i) \emph{learning target} and available inputs/feature \emph{data}; (ii) ML model \emph{hypothesis class} and \emph{training loss}; (iii) \emph{training algorithm} for optimizing ML parameters; and (iv) \emph{evaluation metrics}.

For standalone ML design, ML acts primarily as a modeling tool for power-system target. At the model selection stage, the model hypothesis is determined by the types of learning target, data pattern, and predictive accuracy metrics. For example, in renewable generation forecasting, recurrent or temporal models may be chosen to capture time-series dynamics, and predictive losses such as mean squared error (MSE) and mean absolute percentage error (MAPE) are used. For stability assessment, the learning target is a system-level stability label or index; convolutional, graph-based, or physics-informed architectures may be employed to exploit spatial structure, and performance is assessed using classification metrics such as F1-score. In practice, this step defines a \emph{candidate set} of model families and hyperparameter configurations to be evaluated. 

The training algorithm is determined by the hypothesis class of $\bm{v}(\cdot;\bm{\theta})$ and training loss. For statistical models (e.g., linear models, support vector machines), $\bm{v}(\cdot;\bm{\theta})$ can often be optimized via convex optimization or other mathematical programming techniques \cite{deisenroth2020mathematics}. For modern deep learning with NN, SGD-based methods are typically used \cite{lecun2015deep}. 

Finally, the standalone ML pipeline is implemented as a nested procedure: an inner loop trains $\bm{\theta}$ for each candidate configuration, while an outer validation loop selects the final model class, hyperparameters, and checkpoint $\bm{\theta}^\star$ based on predictive metrics and generalization on a held-out validation set. As training is nonconvex and stochastic (e.g., due to initialization, mini-batching, and early stopping, etc), $\bm{\theta}^\star$ is not guaranteed to be a global minimizer.

\subsection{Integrated LAPSO Pipeline}\label{sec:lapso_framework_lapso}

The LAPSO as an integrated ML design framework follows the same workflow as standalone ML. However, as the learned parameter $\bm{\theta}$ parameterizes the downstream optimization problem through the learned component (e.g., via learning-dependent objective terms $f_v(\cdot)$ and/or auxiliary constraints $\bm g_v(\cdot)\le\tau$), LAPSO also requires specifying (v) \emph{class of PSO}, beyond the four existing primitives (Fig.~\ref{fig:learning_pipeline}). The class $\mathcal{P}_{basic}$ captures the underlying structure of PSO problems, e.g. UC as an MILP, ED as an LP, and OPF as nonlinear and nonconvex programs, etc.

\subsubsection{Optimization-aware Metrics}
In addition to standalone predictive metrics, LAPSO evaluates the optimization layer using four general \emph{optimization-aware metric} axes, which are intended to be application-independent but their concrete realization depends on the operational setting:
\begin{itemize}[leftmargin=*]
    \item \textbf{Solution quality indicator.} 
    Solution quality is assessed by the strength and guarantees provided by the solver and class of PSO, including global optimality guarantees for convex problems or quantified optimality gaps for mixed-integer formulations.
    \item \textbf{Computational tractability.} 
    The deployed problem $\mathcal{P}_{lapso}$ should be solved reliably within the operational time budget, given the solver and the optimization class of $\mathcal{P}_{basic}$. In practice, tractability is assessed via runtime, scalability with problem size, and solver robustness.
    \item \textbf{Constraint satisfaction.} 
    The deployed solution should satisfy physical and operational constraints. When soft constraints or relaxations are used, constraint satisfaction is assessed via the frequency and severity of violations.
    \item \textbf{Economic performance.} 
    System-level performance is assessed via the objective of $\mathcal{P}_{basic}$ or the realized end-to-end operational cost. Comparisons should be made under matched operational requirements when feasible sets differ.
\end{itemize}

\subsubsection{Model Selection}

The model hypothesis is determined by the types of learning target, data pattern, ML and optimization-aware metrics. At deployment, the learned parameters $\bm\theta^\star$ are fixed, while the operational decision variables $\bm z$ are optimized online. 
We summarize the main implications for selecting model as follows.

\begin{itemize}[leftmargin=*]
    \item \textbf{Solution-quality indicator and computational tractability.} 
    $\mathcal{P}_{lapso}$ typically preserves or increases the modeling and computational burden relative to $\mathcal{P}_{basic}$. The hypothesis class selected for $\bm{v}(\cdot)$ determines the analytical form of the learning-dependent terms, e.g., $\bm g_v(\cdot)\leq\bm\tau$ and/or $f_v(\cdot)$, and how they can be represented within the downstream solver. As a result, the impact of the learned component on the complexity depends on (i) whether the learned component depends explicitly on the optimization variable and (ii) how it is integrated into the optimization layer. For the \emph{decision-dependent} case, the functional class of $\bm{v}(\cdot;\bm{\theta}^\star)$ and its encoding can materially change the induced problem size and even the optimization class, thereby affecting both computational tractability and the strength of available solution-quality guarantees. For example, piecewise-linear NN encodings may introduce auxiliary and binary variables, whereas input-convex neural networks (ICNNs) can preserve convexity under suitable representations. For a \emph{decision-independent} predictive model, evaluating $\bm{v}(\cdot;\bm{\theta}^\star)$ often only instantiates numerical quantities in $\mathcal{P}_{lapso}$ and therefore can likely preserve the original optimization class. 
    \item \textbf{Predictive accuracy and constraint satisfaction.} When a learned constraint $\bm{g}_v(\cdot)\leq\tau$ is introduced, predictive performance of $\bm{v}(\cdot)$ becomes directly linked to constraint satisfaction: model mismatch translates into increased probability of violating the intended stability requirement (e.g. false negatives) during deployment, even if the surrogate constraint is feasible within the optimization model.
    \item \textbf{Economic performance.} Model should be selected to balance \emph{economic performance} against operational stability requirements. In particular, introducing a learned constraint $\bm{g}_v(\cdot)\leq\tau$ induces a clear trade-off.
\end{itemize}

\subsubsection{Training}

The training algorithm is determined by the hypothesis class, training loss and the class of PSO. Incorporating the optimization layer into training generally modifies the standard ML pipeline, as system-level objectives depend on $\bm{\theta}$ through the downstream operational problem. Therefore, training complexity depends not only on the ML model itself, but also on how the optimization layer is handled, for example through mathematical-program-based reformulations or gradient-based methods such as differentiable optimization. Even when deployment-time inference is simple, training can still be computationally demanding if it requires repeated solution and differentiation of embedded optimization problems, or repeated optimization-based evaluation during data generation and validation.

\subsubsection{Validation} 

When selecting $\bm{\theta}^\star$ (and associated design choices such as $\tau$ and hyperparameters), validation is considered as an outer loop procedure, driven by the optimization-aware model-selection mechanism in addition to the predictive accuracy. This is because predictive improvements need not translate monotonically into improved system-level outcomes in $\mathcal{P}_{lapso}$.

Accordingly, a key design goal in LAPSO is to incorporate learning modules in a way that \emph{preserves, as much as possible,} the favorable structure of $\mathcal{P}_{basic}$ without incurring excessive cost; when this is not possible, the trade-offs in tractability and solution quality should be made explicit and quantified.

\begin{table*}[t]
\centering
\footnotesize
\renewcommand{\arraystretch}{1.15}
\caption{Summary of standalone ML and various LAPSO pipelines over different coupling patterns.}
\label{tab:lapso_coupling_guidance}
\begin{tabularx}{\linewidth}{p{2.6cm}p{3.8cm}p{3.3cm}X}
\toprule
\textbf{Coupling pattern} & \textbf{How it enters $\mathcal{P}_{basic}$} & \textbf{Deployment-time impact} & \textbf{Model-selection rule} \\
\midrule
\textbf{Standalone ML} \newline Section~\ref{sec:lapso_framework_ml}
& No optimization embedding.
& Only forward evaluation of the learned model.
& Among the candidates, select the model class, hyperparameters, and checkpoint, etc using predictive metrics on validation set. \\
\midrule
\textbf{Decision-dependent predictor} \newline Example: SCO \newline Section~\ref{sec:lapso_framework_lapso} and \ref{sec:special_sco}
& Embedded as an auxiliary constraint, e.g., $u(\bm{z},\bm{y};\bm{\theta}_u^\star)\le\tau$, to $\mathcal{P}_{basic}$.
& May change the class of optimization; affects all optimization-aware metrics of $\mathcal{P}_{basic}$.
& For candidate assessor, first enforce operational acceptability on $\mathcal{P}_{sco}$, such as solve time and stability performance, then select the model that best balances realized cost and conservatism (compared to $\mathcal{P}_{basic}$). \\
\midrule
\textbf{Decision-independent predictor} \newline Example: OBF \newline Section~\ref{sec:lapso_framework_lapso} and \ref{sec:special_obf}
& A predicted parameter $\hat{\bm{y}}_2=\bm{h}(\bm{x};\bm{\theta}_h^\star)$ does not depend on $\bm{z}$ and parameterizes $\mathcal{P}_{basic}$ without adding learned constraints.
& Optimization class is typically unchanged; runtime is dominated by $\mathcal{P}_{basic}$ plus forward evaluation of $\bm{h}(\cdot)$.
& Select the forecaster model class, training loss, and training algorithm by embedding $\hat{\bm{y}}_2$ into the operational problem or decision chain and validating optimization-aware metrics mainly on the realized cost. \\
\midrule
\textbf{Closed-chain hybrid predictor} \newline Example: OBF--SCO \newline Section~\ref{sec:lapso_framework_lapso} and \ref{sec:obf-sco}
& A forecast $\hat{\bm{y}}_2=\bm{h}(\bm{x};\bm{\theta}_h^\star)$ parameterizes both $\mathcal{P}_{basic}$ and decision-dependent $u(\bm{z},\bm{y}_1,\hat{\bm{y}}_2;\bm{\theta}_u^\star)\le\tau$.
& May change the class of optimization; affects all optimization-aware metrics of $\mathcal{P}_{basic}$.
& Select the forecaster model class, training loss, and training algorithm by embedding $\hat{\bm{y}}_2$ into the stability-constrained operational chain as $\mathcal{P}_{obf/sco}$ and validate jointly on all optimization-aware metrics. \\
\bottomrule
\end{tabularx}
\end{table*}

\begin{remark}[Clarification on the terminologies and existing literature]
    The definition of LAPSO covers the decision-focused learning (DFL) and constraint learning in literature. Many classical DFL formulations \cite{beichter2024decision, zhang2025decision,vohra2023end,wahdany2023more}, including beyond power system applications \cite{mandi2024decision}, focus on the decision-independent predictive model such as OBF, which is also referred as cost (or value)-oriented forecasting \cite{zhang2022cost}, task-aware learning \cite{xu2024task} and predict-and-optimize \cite{xie2024predict}. Constraint learning \cite{maragno2025mixed,zhang2023data} acquires the underlying rules, boundaries, or restrictions of a system from data for objective or constraints in an optimization that are not explicitly known. LAPSO covers both research directions (Fig.~\ref{fig:lapso_scope}) by quantifying their difference as decision-independent and decision-dependent predictive models. It moves beyond unified formulation by proposing practical design pipeline under tighter learning-optimization interaction in power systems. 
\end{remark}

\section{Representative Instantiations of LAPSO}\label{sec:special_type}

This section presents SCO (as decision-dependent predictor) and OBF (as decision-independent predictor) as representative examples to demonstrate how LAPSO unifies their mathematical formulation and how it can be used for optimization-aware ML design. The standalone ML and LAPSO pipelines of different learning-optimization coupling patterns are summarized in Table~\ref{tab:lapso_coupling_guidance}.

To avoid ambiguity, we use $\mathcal{P}$ to denote a deployment-time power-system optimization problem $\mathcal{P}_{basic}$ or $\mathcal{P}_{lapso}$ whose decision variable is the operational decision $\bm{z}$. We use $\mathcal{M}$ to denote the ML training problem whose decision variable is the model parameter $\bm{\theta}$. In the LAPSO framework, $\mathcal{M}$ may depend on $\mathcal{P}$ through an implicit or nested structure, since system-level training signals are defined via the optimal solution and/or value of $\mathcal{P}$.

\subsection{Stability-Constrained Optimization}\label{sec:special_sco}

The goal of SCO is to incorporate a learned stability surrogate into an operational optimization so that the resulting decisions satisfy stability requirements while maintaining acceptable computational tractability and economic performance.

\subsubsection{LAPSO Primitives and Formulation}
\label{sec:sco_primitives}

Following Fig.~\ref{fig:learning_pipeline}, SCO instantiates the LAPSO primitives as follows.
(i) \emph{learning target:} a stability label or index $\xi$ e.g. small-signal stability, defined on operating points;
(ii) \emph{hypothesis class:} ML model $\bm{v}(\bm{z},\bm{y}_1,\bm{x};\bm{\theta}^\star)$ is simplified as $u(\bm{z},\bm{y};\bm{\theta}_u^\star)\in\mathbb{R}$, which maps from decision variables $\bm{z}$ and parameters $\bm{y}$ into the stability index ${\xi}$;
(iii) \emph{metrics:} component-level predictive metrics and \emph{optimization-aware} system metrics for model selection; and
(iv) \emph{training algorithm:} standard supervised learning with MP or SGD for fitting $u(\cdot)$;
(v) \emph{class of PSO:} a baseline operational problem $\mathcal{P}_{basic}$.

The supervised training loss is written as
\begin{equation*}
    \boxed{\mathcal{M}_{sco}}:\quad \min_{\bm{\theta}_u}~~ \frac{1}{|\mathcal{D}|} \sum_{((\bm{z},\bm{y}),\xi) \in\mathcal{D}} -H(\xi,u(\bm{z},\bm{y};\bm{\theta}_u)),
\end{equation*}
where $H$ is the binary cross-entropy; $\{(\bm{z},\bm{y}),\xi\}\in\mathcal{D}$ is a sampled data; $\xi$ is the $\{0,1\}$ label where $0$ represents a stable sample. Because $\mathcal{M}_{sco}$ does not interact with the $\mathcal{P}_{basic}$ during training, it is referred to as \emph{accuracy-based} or \emph{open-loop} training. 
Once it is trained, $\mathcal{P}_{lapso}$ becomes,
\begin{equation*}
    \boxed{\mathcal{P}_{sco}}:\quad \min_{\bm{z}} ~ f(\bm{z},\bm y)
        \quad \text{s.t.} ~ \bm{g}(\bm{z};\bm{y}) \leq \bm{0}, ~u(\bm{z},\bm{y};\bm{\theta}^\star_u)\leq 0,
\end{equation*}
i.e., a new stability constraint $u(\cdot;\bm{\theta}_u^\star)$ is added to $\mathcal{P}_{basic}$. It is straightforward to see that $\mathcal{P}_{sco}$ is a special type of $\mathcal{P}_{lapso}$ by setting $\bm{y}_1:=\bm{y}$, $\tau=0$, and $g_v(\bm{z};\bm{y}_1,\hat{\bm y}_2) =  \hat{\bm{y}}_2 = u(\bm{z},\bm{y};\bm{\theta}^\star_u)$.

\subsubsection{Model Selection and Validation}
\label{sec:sco_validation}

In SCO, the stability assessor is both (i) an ML predictor and (ii) a constraint-generating module in the downstream optimization. Therefore, model selection and validation must incorporate \emph{optimization-aware} system metrics, cf. Fig.~\ref{fig:learning_pipeline}.  

\begin{figure}[t]
     \centering
     \begin{subfigure}[b]{0.18\textwidth}
         \centering
         \includegraphics[width=\textwidth]{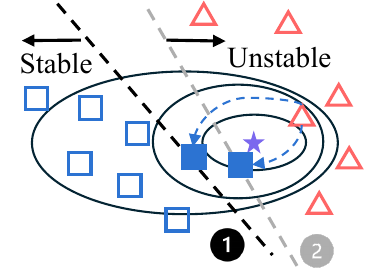}
         \caption{}
         \label{fig:sco_design_a}
     \end{subfigure}
     \hspace{0.06\linewidth}
     \begin{subfigure}[b]{0.18\textwidth}
         \centering
         \includegraphics[width=\textwidth]{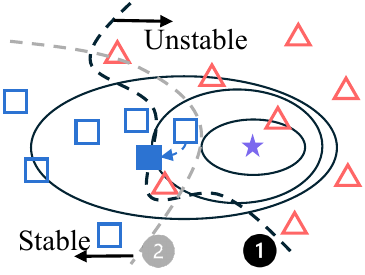}
         \caption{}
         \label{fig:sco_design_b}
     \end{subfigure}
        \caption{Illustrative examples on the trade-off between predictive accuracy/constraint satisfaction and economic performance in SCO. The purple star represents the minimizer of $\mathcal{P}_{basic}$'s objective; Blue and red points represent stable and unstable samples; Hollow and solid points represent the sample with and without stability constraints. The stability boundary is shown by dashed lines/curves. In (a), both assessors \textcircled{1} and \textcircled{2} match the labeled samples equally well and achieve 100\% accuracy, yet impose different constraint geometries in the operating space, leading to different optimal operating points under $\mathcal{P}_{sco}$. For the highlighted unstable sample, assessor \textcircled{2} enables a more economical operation as the boundary is closer to the optimum of $\mathcal{P}_{basic}$'s objective; In (b), a conservative stability boundary \textcircled{1} correctly identifies all unstable samples but misclassifies one stable sample as unstable (false positive sample), thereby increasing its operational cost after the unnecessary correction. In contrast, a simpler boundary \textcircled{2} does not accurately assess all unstable samples but all stable samples are operated optimally under $\mathcal{P}_{sco}$ as $\mathcal{P}_{basic}$.}
        \label{fig:sco_design}
\end{figure}

\paragraph{Computational tractability and solution quality indicator} Referring to Section~\ref{sec:lapso_framework_lapso}, SCO learns a \emph{decision-dependent model} such that its encoding can change the original class of $\mathcal{P}_{basic}$, affecting the convergence and computational burden of the solution algorithm. In detail, if the original problem is an MILP, such as UC, the stability constraint should ideally be a convex function (such as linear or second-order cone) or at least can be reformulated as a mixed integer linear (MIL) function after encoding such that $\mathcal{P}_{sco}$ has the same optimization class as $\mathcal{P}_{basic}$. This setting ensures the computational tractability and solution quality, which has been explored in previous research \cite{chu2023stability,xu2024incorporation,cui2025control, jia2025converter} by training a convex assessor. Otherwise, the global optimality of $\mathcal{P}_{sco}$ may not be achieved in which explicit approximation and iterative algorithm is required \cite{wang2025synchronous}.

\paragraph{Predictive accuracy, constraint satisfaction, and economic performance} When ML in $\mathcal{M}_{sco}$ is considered as a modeling tool, its complexity determines the accuracy, which will eventually be reflected in real-time grid stability performance. We state that convex constraint is preferable for computational tractability and solution quality; however, since the actual stability criterion is unlikely a convex function of $\bm{z}$ and $\bm{y}$ nor readily computable without solving ad-hoc differential equations, there will inevitably be unstable samples that remain undetected. 

In addition, a key reason predictive accuracy is insufficient in SCO is that the embedded constraint depends on the \emph{geometry} of the learned stability surrogate over the continuous operating region explored by the optimizer, not only on correctness at a finite set of labeled samples. Such geometry will inevitably influence the operational cost under deployment. 

In detail, Fig.~\ref{fig:sco_design_a} illustrates that two assessors can achieve the same (or even perfect) classification accuracy on a finite dataset, yet still induce different feasible sets when enforced as a constraint through $u(\bm z,\bm y;\bm\theta_u^\star)\leq 0$, due to differences in boundary shape between sampled points and/or differences in stability margins. Since the optimizer selects $\bm z^\star$ by trading cost against feasibility of $\mathcal{P}_{sco}$, these geometric differences can lead to different operational solutions and different realized costs. In addition, Fig.~\ref{fig:sco_design_b} highlights that conservative surrogate boundaries can increase operational cost by misclassifying safe points, while less conservative boundaries may miss a limited number of unstable cases. 

We contend that pursuing perfect accuracy with conservative assessor for rare events at the expense of higher computational burden and degraded economic efficiency at the operation stage is unnecessary. Within a sequential decision-making framework, a limited number of misclassifications can still be corrected at the control stage. The SCO pipeline is summarized in Table~\ref{tab:lapso_coupling_guidance}.

\subsection{Objective-based Forecasting}
\label{sec:special_obf}

This section instantiates the integrated LAPSO pipeline in Section~\ref{sec:lapso_framework_lapso} on OBF. The goal of OBF is to design and validate an energy forecaster so that its predictions lead to improved system-level operational outcomes when embedded in downstream PSO problem(s).

\subsubsection{LAPSO Primitives and Formulation}
\label{sec:obf_primitives}

Following Fig.~\ref{fig:learning_pipeline}, OBF instantiates the LAPSO design primitives as:
(i) \emph{learning target:} a predictable trajectory $\bm{y}_2$ such as renewable generation or load;
(ii) \emph{hypothesis class:} a forecaster $\hat{\bm{y}}_2=\bm{h}(\bm{x};\bm{\theta}_h)$ mapping contextual features $\bm{x}$ such as historical profiles, weather, calendar information to $\hat{\bm{y}}_2$;
(iii) \emph{metrics:} component-level predictive metrics \emph{and} optimization-aware metrics;
(iv) \emph{training algorithm:} methods that account for the dependence of system-level objectives on $\bm{\theta}_h$ through mathematical program or gradient-based algorithms; and
(v) \emph{Class of PSO:} an operational optimization $\mathcal{P}_{basic}$.

The OBF problem during deployment specializes $\mathcal{P}_{lapso}$ as,
\begin{equation*}
    \begin{aligned}
       \boxed{\mathcal{P}_{obf}}:\quad \min_{\bm{z}}~~  & f(\bm{z},\bm{y}_1) \\
        \text{s.t.}~~ & \bm{g}_1(\bm{z}, \bm{y}_1) \leq \bm{g}_2(\bm{y}_1,\hat{\bm{y}}_2) \\
        & \hat{\bm{y}}_2 = \bm{h}(\bm{x};\bm{\theta}^\star_h), 
    \end{aligned}
\end{equation*}
where $\bm{y}_1$ collects quantities treated as known/exogenous at decision time. $\mathcal{P}_{obf}$ can be obtained from $\mathcal{P}_{lapso}$ by setting $\hat{\bm y}_2=\bm{v}(\bm{z},\bm{y}_1,\bm{x};\bm{\theta}^\star) = \bm{h}(\bm{x};\bm{\theta}_h^\star)$ and $\bm{g}(\bm{z};\bm{y}_1,\hat{\bm y}_2) = \bm{g}_1(\bm{z},\bm{y}_1) - \bm{g}_2(\bm{y}_1,\hat{\bm y}_2) \leq \bm{0}$. Compared with the SCO setting, the coupling between $\bm{z}$ and $\hat{\bm{y}}_2$ is through the parameterization of constraints (here represented by $\bm{g}_2$), and $\bm{\tau}$ is not required. Similarly to the SCO, a standalone baseline trains $\bm{h}(\cdot;\bm{\theta}_h)$ by minimizing MSE, defined as \emph{accuracy-based forecast} (ABF):
\begin{equation*}
    \boxed{\mathcal{M}_{abf}}:\quad \min_{\bm{\theta}_h} \frac{1}{|\mathcal{D}|} \sum_{(\bm{x}, \bm{y}_2)\in\mathcal{D}} \|\bm{y}_2 - \bm{h}(\bm{x};\bm{\theta}_h) \|_2^2.
\end{equation*}

\subsubsection{Training Loss}

In LAPSO, the training objective is defined using \emph{optimization-aware} system outcomes induced by solving the operational problem $\mathcal{P}_{obf}$. It should therefore respect both the operational objective(s) and the information timing of the decision chain induced by $\mathcal{P}_{obf}$. We summarize two representative schemes below. 

First, we summarize previous works \cite{munoz2022bilevel, chen2024towards, xu2024e2e, vohra2023end, beichter2024decision, xu2024task} as \emph{closed-loop OBF} training, which evaluates the realized end-to-end operational outcome by executing the same decision chain as in deployment. Taking a two-stage power-system decision-making process as an example, such as UC / DP followed by ED / RD, closed-loop OBF can be written as the following bi-level optimization:
\begin{equation*}
    \begin{aligned}
        & \boxed{\mathcal{M}_{obf}}: ~ \min_{\bm{\theta}_h} \sum_{(\bm{x},\bm{y}_1, \bm{y}_2)\in\mathcal{D}} \ell(\hat{\bm{z}}^2, \hat{\bm{z}}^1, \bm{y}_1) \\
        \text{s.t.} ~& \text{For } \forall (\bm{x},\bm{y}_1,\bm{y}_2)\in\mathcal{D}, \\
            & \hspace{0.3cm} \hat{\bm{z}}^2 = \arg\min_{\bm{z}^2} \left\{f^2(\bm{z}^2,\bm{y}_1): \bm{g}_1^2(\bm{z}^2,\bm{y}_1) \leq \bm{g}_2^2(\bm{y}_1,\hat{\bm{z}}^1, \bm{y}_2)\right\} \\
            & \hspace{0.3cm} \hat{\bm{z}}^1 = \arg\min_{\bm{z}^1} \left\{f^1(\bm{z}^1,\bm{y}_1): \bm{g}_1^1(\bm{z}^1,\bm{y}_1) \leq \bm{g}_2^1(\bm{y}_1,\hat{\bm{y}}_2)\right\} \\
        & \hspace{0.3cm} \hat{\bm{y}}_2 = \bm{h}(\bm{x};\bm{\theta}_h),
    \end{aligned}
\end{equation*}
where the superscripts $(\cdot)^1$ and $(\cdot)^2$ denote the stage-1 and stage-2 components, respectively. In this setting, the operational problems induced by $\mathcal{P}_{obf}$ appear as the lower-level problems, and the training loss $\ell(\cdot)$ is defined using the realized end-to-end outcome on realized quantities, consistent with real-time deployment. From a game-theoretic perspective, $\mathcal{M}_{obf}$ can be interpreted as a leader--follower Stackelberg game, where the leader optimizes the shared forecaster parameter $\bm{\theta}_h$, and the followers solve the downstream operational problems for each sample using the induced forecast $\hat{\bm{y}}_2$. The above formulation is sufficiently general to cover the DP / RD setting used in the case study in Section~\ref{sec:obf_case_study}; for other applications, the exact structure may vary, but the underlying principle remains the same.

Second, previous works \cite{chen2021feature, wahdany2023more, smetsdecision2024decision} are summarized as \emph{counterfactual OBF} training, which first computes an \emph{oracle} target or optimal decision using realized $\bm{y}_2$ for each sample in the dataset. During training, the loss is then defined as a supervised discrepancy to this oracle, which is often referred to as a \emph{regret}-type loss:
\begin{equation*}
    \begin{aligned}
        \min_{\bm{\theta}_h} \; & \sum_{(\bm{x},\bm{y}_1,\bm{y}_2)\in\mathcal{D}}
        \left(f^1(\hat{\bm{z}}^1,\bm{y}_1) - f^1(\bm{z}^{1,\star},\bm{y}_1)\right)^2 \\
        \text{s.t.}\quad 
        & \text{For } \forall (\bm{x},\bm{y}_1,\bm{y}_2)\in\mathcal{D}, \\
        & \hspace{0.3cm} \hat{\bm{z}}^1 = \arg\min_{\bm{z}^1} \left\{ f^1(\bm{z}^1,\bm{y}_1) :
        \bm{g}_1^1(\bm{z}^1,\bm{y}_1) \leq \bm{g}_2^1(\bm{y}_1,\hat{\bm{y}}_2) \right\} \\
        & \hspace{0.3cm} \hat{\bm{y}}_2 = \bm{h}(\bm{x};\bm{\theta}_h),
    \end{aligned}
\end{equation*}
where $f^1(\bm{z}^{1,\star},\bm{y}_1)$ denotes the optimal value of the stage-1 problem solved using the realized target $\bm{y}_2$, which is unavailable at deployment time. Compared with closed-loop $\mathcal{M}_{obf}$, only the stage-1 operational problem is embedded in the lower level.

\begin{figure}[t]
    \centering
    \includegraphics[width=0.99\linewidth, trim={0.0cm 0.0cm 0.0cm 0.0cm}, clip]{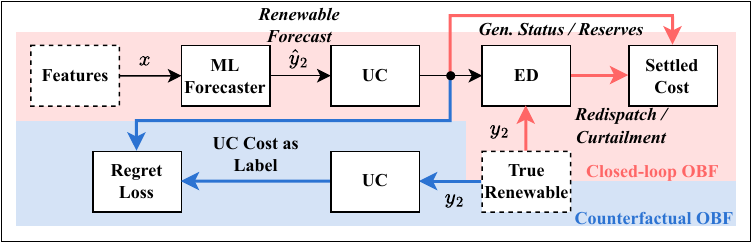}
    \caption{UC--ED example illustrating OBF training loss regimes. In the closed-loop OBF, the full forecast--UC--ED decision chain is incorporated into training, allowing the loss to be defined by the realized end-to-end operational cost. In contrast, the counterfactual setting includes only the UC stage and relies on an \emph{offline labeling process} to compute the oracle UC decision and cost driven by the realized renewable profile $\bm{y}_2$.}
    \label{fig:obf_framework}
\end{figure}

The UC-ED example is illustrated in Fig.~\ref{fig:obf_framework}. Note that both training regimes use realized information $\bm{y}_2$ during training; the main difference lies in the timing and interpretation of the supervision signal.

\subsubsection{Model Selection and Training}

Solving $\mathcal{M}_{obf}$ can be challenging, depending on \emph{hypothesis class} of the ML model and the \emph{class of PSO}. Some representative examples are summarized in Table \ref{tab:obf_train} with different optimality guarantees on $\bm{\theta}_h$. Notably, for convex $\mathcal{P}_{basic}$, it is possible to denote the lower-level problem(s) as mathematical program(s) with equilibrium constraints (MPEC) through the Karush-Kuhn-Tucker (KKT) conditions at optimality \cite{gabriel2012complementarity}, which allows a direct solution by calling optimization solver if the upper level of $\mathcal{M}_{obf}$ is also (mixed-integer) convex. Apparently, this requires the forecasting model to be also convex, such as linear regression (LR). In contrast, modern deep learning requires SGD to efficiently update the NN parameters via automatic differentiation (AD) technique \cite{lecun2015deep}. As a result, implicit differentiation (ID) or differentiable optimization \cite{blondel2024elements} is needed to find the gradient through the optimization problem in the backward pass after the exact $\mathcal{P}_{basic}$ is solved in the forward pass. 

\begin{table}[t]
    \centering
    \footnotesize
    \renewcommand{\arraystretch}{1.2} 
    \caption{Examples training algorithms in $\mathcal{M}_{obf}$.}
    \begin{threeparttable}
    \begin{tabularx}{\linewidth}{cXp{4.1cm}p{0.9cm}}
        \toprule
        \textbf{Ref.} & $\mathcal{P}_{basic}$/$\bm{h}(\cdot;\bm{\theta}_h)$ & \textbf{Typical Solution Approach} & \textbf{Global Optim.} \\
        \midrule
        \cite{munoz2022bilevel} & Convex / Convex & Optimization solver: transform $\mathcal{M}_{obf}$ into MPEC & \cmark$^\diamond$ \\
        \midrule
        \cite{xu2024e2e} & Convex / NN & Hybrid SGD + optimization solver: AD through $\bm{h}(\cdot;\bm{\theta}_h)$ and ID via KKT conditions of $\mathcal{P}_{basic}$ & \xmark \\
        \midrule
        \cite{chen2024towards} & Mixed-integer Convex / Convex & Optimization solver: transform $\mathcal{M}_{obf}$ into MPEC and solve by column-and-constraint generation (C\&CG) & \cmark$^\diamond$ \\
        \midrule
        \cite{zhou2024load} & Mixed-integer Convex / NN & Hybrid SGD + optimization solver: AD through $\bm{h}(\cdot;\bm{\theta}_h)$ and ID via KKT conditions of $\mathcal{P}_{basic}$ (with relaxed integer variables) & \xmark \\
        \midrule
        \cite{kotary2023backpropagation} & Nonconvex / NN & Hybrid SGD + optimization solver: AD via $\bm{h}(\cdot;\bm{\theta}_h)$ and unrolled $\mathcal{P}_{basic}$ & \xmark \\
        \bottomrule
    \end{tabularx}
    \begin{tablenotes}
      \item[$\diamond$] Subject to a predefined convergence tolerance gap.  
      \end{tablenotes}
    \end{threeparttable}
    \label{tab:obf_train}
\end{table}

\subsubsection{Validation}
Consistent with Section~\ref{sec:lapso_framework_lapso}, OBF is naturally interpreted as an optimization-aware model validation procedure as follows.

\paragraph{Solution quality indicator and Computational tractability} It is observed that the decision $\bm{z}$ and the forecaster $\hat{\bm{y}}_2$ are disentangled in $\mathcal{P}_{obf}$, which follows the decision-independent predictive model setting. As discussed in Section~\ref{sec:lapso_framework_lapso}, the disentanglement suggests that, in the development stage, $\mathcal{P}_{obf}$ can first take forecast $\hat{\bm{y}}_2=\bm{h}(\bm{x};\bm{\theta}^\star_h)$ and then compute $\bm{g}_2(\bm{y}_1,\hat{\bm{y}}_2)$ into optimization $\mathcal{P}_{basic}$, which uncovers the exact decision-making sequence in real time. Therefore, unlike the SCO case, the complexity of the forecaster does not influence the efficiency and solution quality of solving $\mathcal{P}_{obf}$ compared to $\mathcal{P}_{basic}$ in general.

\paragraph{Predictive Accuracy, economic performance, and constraint satisfaction} As defined in previous section, $\mathcal{M}_{obf}$ is trained to align the forecaster with the economic performance such as realized end-to-end cost or regret relative to an oracle benchmark. Unless a perfect forecaster exists, there will be a mismatch between the predictive accuracy and the economic performance. This is mainly triggered by reconfiguration of the constraint set of $\mathcal{P}_{obf}$. 

However, unlike the standalone ML-based forecaster, the $\mathcal{M}_{obf}$ is directly associated with the specific choice of $\mathcal{P}_{basic}$s, which demonstrates another layer of uncertainties if $\mathcal{P}_{basic}$ at the deployment stage is different from what is considered during training. An immediate example is that the different choices of training regimes (closed-loop or counterfactual OBFs) will result in different forecaster parameters, which cannot be generalized between each other \cite{xu2024e2e}. Further discussion is presented in Section~\ref{sec:obf-uncertain} on uncertainty quantification in LAPSO.  

The LAPSO pipeline of OBF is summarized in Table~\ref{tab:lapso_coupling_guidance}.

\section{Closed-loop Integration and Uncertainty Quantification}\label{sec:new_application}

\subsection{Integrated Forecast--Operation--Control}
\label{sec:obf-sco}

A key advantage of LAPSO is that the same optimization-centered template can be extended beyond single-task settings, such as SCO-only or OBF-only formulations. Here we illustrate this extensibility by \emph{closing the forecast--operation--control chain}: the forecast parameterizes the operations problem, stability requirements enter as decision-dependent constraints, and the resulting system-level outcomes are used to train the forecaster.

\subsubsection{Formulation}

Following OBF construction, we replace $\mathcal{P}_{basic}$ with the stability-constrained operation layer, cf. Fig.~\ref{fig:lapso_scope}, yielding the following deployment-time hybrid problem:
\begin{equation*}
\begin{aligned}
\boxed{\mathcal{P}_{obf/sco}}:\; \min_{\bm{z}}~~ & f(\bm{z},\bm{y}_1) \\
\text{s.t.}~~ & \bm{g}_1(\bm{z},\bm{y}_1) \leq \bm{g}_2(\bm{y}_1,\hat{\bm{y}}_2) \\
& u(\bm{z},\bm{y}_1,\hat{\bm{y}}_2;\bm{\theta}_u^\star) \leq 0 \\
& \hat{\bm{y}}_2 = \bm{h}(\bm{x};\bm{\theta}_h^\star),
\end{aligned}
\end{equation*}
where the predicted trajectory $\hat{\bm{y}}_2$ enters both the physics-based constraint parameterization $\bm{g}_2(\cdot)$ and the stability constraint $u(\cdot)$, and $\bm{\theta}_u^\star$ denotes a pre-trained stability assessor.

In principle, one may jointly learn $(\bm{\theta}_u,\bm{\theta}_h)$. To keep the exposition focused, we consider a simplified setting in which the stability assessor is trained offline via $\mathcal{M}_{sco}$ and then kept fixed when training the forecaster. Adapting the closed-loop OBF bi-level template accordingly yields
\begin{equation*}
    \begin{aligned}
        & \boxed{\mathcal{M}_{obf/sco}}: ~ \min_{\bm{\theta}_h} \sum_{(\bm{x},\bm{y}_1, \bm{y}_2)\in\mathcal{D}} \ell(\hat{\bm{z}}^2, \hat{\bm{z}}^1, \bm{y}_1) \\
        \text{s.t.} ~~& \text{For } \forall (\bm{x},\bm{y}_1,\bm{y}_2)\in\mathcal{D}, \\
            & \hspace{0.3cm} \hat{\bm{z}}^2 = \arg\min_{\bm{z}^2} \{f^2(\bm{z}^2,\bm{y}_1):  u(\bm{z}^2, \bm{y}_1, \bm{y}_2;\bm{\theta}_u^\star) \leq 0, \\
            & \hspace{4.0cm} \bm{g}_1^2(\bm{z}^2,\bm{y}_1) \leq \bm{g}_2^2(\bm{y}_1,\hat{\bm{z}}^1, \bm{y}_2) \} \\
            & \hspace{0.3cm} \hat{\bm{z}}^1 = \arg\min_{\bm{z}^1} \{f^1(\bm{z}^1,\bm{y}_1): u(\bm{z}^1, \bm{y}_1, \hat{\bm{y}}_2;\bm{\theta}_u^\star) \leq 0,\\
            & \hspace{4.0cm} \bm{g}_1^1(\bm{z}^1,\bm{y}_1) \leq \bm{g}_2^1(\bm{y}_1,\hat{\bm{y}}_2)  \}
         \\
        & \hspace{0.3cm} \hat{\bm{y}}_2 = \bm{h}(\bm{x};\bm{\theta}_h),
    \end{aligned}
\end{equation*}
that is, the lower-level operational problems in closed-loop OBF are replaced by their stability-constrained counterparts under $\mathcal{P}_{obf/sco}$.

\subsubsection{LAPSO Interpretation and Design Implications}

$\mathcal{P}_{obf/sco}$ and $\mathcal{M}_{obf/sco}$ exemplify a hybrid LAPSO setting in which a \emph{decision-independent} predictor $\bm h(\bm x;\bm\theta_h)$ is coupled with a \emph{decision-dependent} learned constraint $u(\bm z,\cdot;\bm\theta_u^\star)$. This hybrid coupling has clear implications under the LAPSO design pipeline (Table~\ref{tab:lapso_coupling_guidance}). First, the forecaster can no longer be evaluated solely by predictive accuracy, because forecast errors propagate not only to \emph{economic performance} through feasible-set reshaping and recourse actions, but also to \emph{constraint satisfaction} through the stability constraint. Second, model selection must account for the \emph{computational tractability} and \emph{solution quality} of the induced lower-level problem, since the addition of a learned stability constraint can alter the optimization class, problem size, and runtime. Third, training requires algorithmic choices that account for the nested dependence of system-level outcomes on $\bm\theta_h$, whether through MP-based reformulations or gradient-based methods (Table~\ref{tab:obf_train}).

\subsection{Uncertainty Quantification}
\label{sec:obf-uncertain}

Previous sections showed that a forecaster can be trained to align with different downstream optimization problems, such as $\mathcal{P}_{obf}$ (Section~\ref{sec:special_obf}) or $\mathcal{P}_{obf/sco}$ (Section~\ref{sec:obf-sco}), and under different OBF regimes defined by information timing such as counterfactual and closed-loop settings; cf. Fig.~\ref{fig:obf_framework}. This section starts by quantifying the source of uncertainties in $\mathcal{P}_{lapso}$ and clarifying how such uncertainty propagates to system-level performance once optimization is in the loop.

\subsubsection{Source of Uncertainty}

Within $\mathcal{P}_{lapso}$, uncertainty can originate from both learning (\emph{ML-Uncertainty}) and the optimization layer (\emph{Opt-Uncertainty}). For clarity, we discuss the OBF setting, where the learned forecaster produces a predictive quantity $\hat{\bm y}_2$ that parameterizes the deployed optimization; the same interpretation extends to the general $\mathcal{P}_{lapso}$ template.

\begin{figure}[t]
    \centering
    \includegraphics[width=0.90\linewidth]{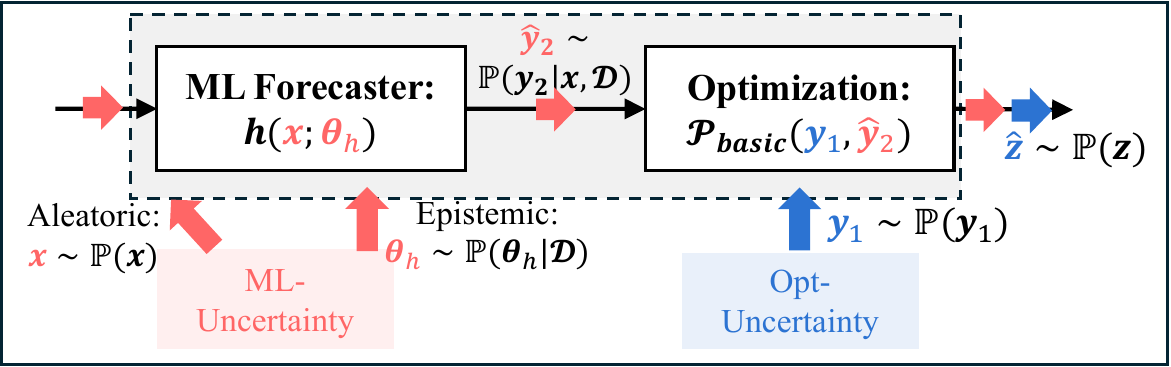}
    \caption{Source and propagation of uncertainties in $\mathcal{P}_{obf}$ for a fixed optimization structure.
    }
    \label{fig:obf-uncertain}
\end{figure}

\paragraph{ML-Uncertainty}
From a Bayesian perspective, uncertainty in the forecaster $\hat{\bm y}_2=\bm h(\bm x;\bm\theta_h^\star)$ can be decomposed into \emph{epistemic} (model) uncertainty and \emph{aleatoric} (data) uncertainty \cite{jospin2022hands}. Epistemic uncertainty captures the forecaster's limited knowledge under finite data and can be represented by a posterior distribution over model parameters, $\mathbb{P}(\bm\theta_h^\star|\mathcal D)$. Aleatoric uncertainty captures inherent randomness in the data and is represented by the likelihood $\mathbb{P}(\bm y_2 | \bm x,\bm\theta_h^\star)$. Marginalizing yields the predictive distribution
\begin{equation*}
\mathbb{P}(\bm y_2| \bm x,\mathcal D)
=
\int \mathbb{P}(\bm y_2| \bm x,\bm\theta_h^\star)\,
\mathbb{P}(\bm\theta_h^\star|\mathcal D)\, d\bm\theta_h^\star.
\end{equation*}
Under LAPSO, the practical relevance of this predictive distribution is that it induces uncertainty in \emph{optimization-aware} system outcomes once $\hat{\bm y}_2$ parameterizes the deployed optimization (Fig.~\ref{fig:obf-uncertain}). 

\paragraph{Optimization-layer uncertainty (Opt-Uncertainty)}

A central step of the LAPSO pipeline (Section~\ref{sec:lapso_framework_lapso}) is validation-driven model selection under optimization-aware metrics. When the downstream PSO structure changes across operational settings (e.g., different $\mathcal{P}_{basic}$ variants), predictive accuracy alone is insufficient to decide whether a forecaster trained under one setting can be reused under another. In this context, the cosine similarity of optimization-induced gradients provides an ex ante diagnostic of \emph{transferability} that fits naturally within the LAPSO design loop.

Consider two operational optimization structures $\mathcal{P}^A$ and $\mathcal{P}^B$ (e.g., two variants of $\mathcal{P}_{basic}$), which induce two system-level training signals through the OBF objective $\ell^{A}$ and $\ell^B$ as in $\mathcal{M}_{obf}$. Also denote the total derivative by $\mathrm{D}(\cdot)$. For a given validation set $\mathcal{D}_{val}$ and a candidate forecaster with parameter $\bm\theta$, compute the corresponding gradients: $\bm\varrho^{A}_i := \mathrm{D}_{\bm\theta}\ell^{A}_i$ and
$\bm\varrho^{B}_i := \mathrm{D}_{\bm\theta}\ell^{B}_i$. Define the sample-wise alignment score
\begin{equation*}
s_i := \frac{\langle \bm\varrho^{A}_i,\bm\varrho^{B}_i\rangle}{\|\bm\varrho^{A}_i\|\,\|\bm\varrho^{B}_i\|}\in[-1,1],
\end{equation*}
and summarize alignment by statistics such as $\bar{s}:=\mathbb{E}_{i\in\mathcal{D}_{val}}[s_i]$. In the LAPSO model-selection and validation stages, gradient-alignment statistics complement predictive metrics and realized system outcomes:
\begin{itemize}[leftmargin=*]
    \item If gradients are consistently aligned (e.g., $\bar{s}$ is high and most $s_i$ are positive), then improvements made by training under $\mathcal{P}^A$ are more likely to also improve economic performance under $\mathcal{P}^B$, supporting reuse/transfer of the forecaster across the two operational settings.
    \item If gradients are frequently misaligned (e.g., many $s_i\le 0$), then training under one structure can degrade performance under the other; in this case retraining is needed.
\end{itemize}
As with other metrics (Fig.~\ref{fig:learning_pipeline}), alignment is a diagnostic rather than a formal guarantee, and should be interpreted under the targeted operational setting.

To compute gradients of the OBF objective with respect to the forecaster parameter, the lower-level optimizations in $\mathcal{M}_{obf}$ must be differentiated. Consider the compact $\mathcal{P}_{lapso}$ formulation. Under standard regularity assumptions (e.g., differentiability and a nonsingular KKT Jacobian of the lower-level problem) at sample $(\bm{x},\bm{y})$, the total derivative of the upper-level objective with respect to $\bm{\theta}$ satisfies
\begin{equation}\label{eq:grad_1}
\bm{\varrho}^{lapso}(\bm{x},\bm{y})
:= \mathrm{D}_{\bm{\theta}} \tilde{f}(\Omega)
= \partial_{\bm{\theta}} \tilde{f}(\Omega)
+ \partial_{\bm{z}} \tilde{f}(\Omega)\cdot \partial_{\bm{\theta}} \hat{\bm{z}}(\bm{\theta}),
\end{equation}
where $\partial_{\divideontimes}(\cdot)$ denotes the partial derivative/Jacobian with respect to $\divideontimes$, and $\Omega = [\hat{\bm{z}},\bm{y}_1,\bm{x},\bm{\theta}]$ denotes the argument list.

Let $\mathcal{KKT}(\Omega)=0$ denote the KKT system of the lower-level optimization defining $\hat{\bm z}(\bm\theta)$. Taking the total derivative of $\mathcal{KKT}(\Omega) = 0$ with respect to $\bm{\theta}$ gives\footnote{Here $\partial_{\bm{z}} \mathcal KKT$ denotes the Jacobian of the KKT system with respect to the primal and dual variables of the lower-level problem.}
\begin{equation*}
\mathrm{D}_{\bm{\theta}} \mathcal{KKT}(\Omega)
= \partial_{\bm{z}}\mathcal{KKT}(\Omega)\cdot \partial_{\bm{\theta}}\hat{\bm{z}}(\bm{\theta})
+ \partial_{\bm{\theta}}\mathcal{KKT}(\Omega)
= 0,
\end{equation*}
and hence the Jacobian of the solution map becomes $\partial_{\bm{\theta}}\hat{\bm{z}}(\bm{\theta})
= -\big[\partial_{\bm{z}}\mathcal{KKT}(\Omega)\big]^{-1}\partial_{\bm{\theta}}\mathcal{KKT}(\Omega)$. Substituting into \eqref{eq:grad_1} yields
\begin{equation}\label{eq:grad}
\begin{aligned}
    & \bm{\varrho}^{lapso}(\bm{x},\bm{y})
    \\ = ~ &\partial_{\bm{\theta}} \tilde{f}(\Omega)
    -  \partial_{\bm{z}} \tilde{f}(\Omega)\cdot
    \big[\partial_{\bm{z}}\mathcal{KKT}(\Omega)\big]^{-1}\partial_{\bm{\theta}}\mathcal{KKT}(\Omega),
\end{aligned}
\end{equation}
which can be evaluated at $\bm{\theta}=\bm{\theta}^\star$ after training.

\subsubsection{Uncertainty-aware Optimization}

When the optimization structure is fixed, uncertainty propagation can be analyzed by separating exogenous inputs $\bm{y}_1$ from predicted quantities $\hat{\bm{y}}_2$. Although $\bm y_1$ is not predicted by the forecaster, it can still be uncertain in practice. For example, an operator responsible for renewable forecasting and operation may face an uncertain load profile provided by a different service provider. To quantify uncertainty in these exogenous optimization inputs, a scenario-based stochastic variant of $\mathcal{P}_{obf}$ can be written as
\begin{equation*}
    \begin{aligned}
       \boxed{\mathcal{P}_{obf/uncer}}:~~ \min_{\bm{z}} ~ & \mathbb{E}_{\Delta \bm{y}_1\sim\mathbb{P}(\Delta \bm{y}_1)}  f(\bm{z},\bm{y}_1 + \Delta \bm y_1) \\
        \text{s.t.} ~ & \bm{g}_1(\bm{z},\bm{y}_1 + \Delta \bm y_1) \leq \bm{g}_2(\bm{y}_1 + \Delta \bm y_1,\hat{\bm{y}}_2) \\
        & \hat{\bm{y}}_2 = \bm{h}(\bm{x};\bm{\theta}^\star_h),
    \end{aligned}
\end{equation*}
where $\Delta \bm y_1$ is the random variable varying against nominal $\bm{y}_1$. Since $\bm y_1$ does not depend on the forecaster, its uncertainty modeling is independent of the ML-uncertainty in $\bm y_2$ at the deployment stage. However, they \emph{jointly affect} optimization-aware metrics through realized cost and constraint-violation behavior by changing the induced optimization instance and the resulting decision. An illustration of these uncertainty sources and their propagation through the LAPSO pipeline is shown in Fig.~\ref{fig:obf-uncertain}.

\subsubsection{Timings of Uncertainty Realization and Uncertainty-aware Training}

Depending on when uncertainty is realized and when operational decisions are made, stochastic/robust optimization typically distinguishes \emph{here-and-now} (recourse decisions made before uncertainty realization) from \emph{wait-and-see} (recourse decisions made after uncertainty realization) settings \cite{beck2023survey}. Many existing works account for ML-uncertainty in a here-and-now manner for predicted parameter, e.g., probabilistic forecasting combined with stochastic dispatch \cite{donti2017task} or analyzing the impact of aleatoric uncertainty on operational outcomes \cite{xu2024e2e, zuo2024transferability}. In these settings, the operational optimization hedges against uncertainty through its own decision variables under the chosen risk model.

We therefore propose a new \emph{wait-and-see} training variant that accounts for uncertainty in \emph{exogenous} parameter $\bm y_1$ by evaluating system-level training signals. Note that such uncertainty is particularly considered in the stage-2 PSO. Concretely, consider the bilevel training problem
\begin{equation*}
    \begin{aligned}
        & \boxed{\mathcal{M}_{obf/uncer}}: \min_{\bm{\theta}_h} \sum_{(\bm{x},\bm{y}_1, \bm{y}_2)\in\mathcal{D}} \mathbb{E}_{\Delta \bm y_1\sim\mathbb{P}(\Delta \bm y_1)} \ell(\hat{\bm{z}}^2, \hat{\bm{z}}^1, \bm{y}_1) \\
        \text{s.t.} ~& \text{For } \forall (\bm{x},\bm{y}_1,\bm{y}_2)\in\mathcal{D}, \\
            & \hspace{0.3cm} \hat{\bm{z}}^2 = \arg\min_{\bm{z}^2} \{f^2(\bm{z}^2,\bm{y}_1 + \Delta \bm y_1): \\
            & \hspace{1.8cm} \bm{g}_1^2(\bm{z}^2,\bm{y}_1 + \Delta \bm y_1) \leq \bm{g}_2^2(\bm{y}_1 + \Delta \bm y_1,\hat{\bm{z}}^1, \bm{y}_2)\} \\
            & \hspace{0.3cm} \hat{\bm{z}}^1 = \arg\min_{\bm{z}^1} \left\{f^1(\bm{z}^1,\bm{y}_1): \bm{g}_1^1(\bm{z}^1,\bm{y}_1) \leq \bm{g}_2^1(\bm{y}_1,\hat{\bm{y}}_2)\right\} \\
        & \hspace{0.3cm} \hat{\bm{y}}_2 = \bm{h}(\bm{x};\bm{\theta}_h),
    \end{aligned}
\end{equation*}
Here the lower-level PSO acts in a wait-and-see manner, while the upper-level selects $\bm{\theta}_h$ by minimizing the expected system-level objective of $\mathcal{M}_{obf}$. Although we present an expectation-based (risk-neutral) form for simplicity, other risk models (e.g., CVaR or robust variants) can be incorporated by replacing the inner expectation with the corresponding risk functional.

\section{Overview of the Lapso Package}\label{sec:package}

Observing the frequent interaction between learning and optimizations in both training and deployment stages, a dedicated Python package \texttt{lapso} is proposed. This package is built upon \texttt{CVXPY} \cite{diamond2016cvxpy} and enables power system optimizations to be formulated in a natural, mathematically intuitive form and solved by the back-end solvers. The \texttt{lapso} package automatically extracts the structures of the \emph{existing} power system optimization problem $\mathcal{P}_{basic}$, along with a neural network model, and compiles them into extended form $\mathcal{P}_{lapso}$, which is then recompiled by \texttt{CVXPY}. The module \texttt{lapso.optimization} supports \emph{mixed-integer parametric quadratic/linear programs}, as denoted as
\begin{equation}\label{eq:standard_integer_qp}
    \begin{aligned}
        \min_{\bm{z}} \; &  \textstyle \frac{1}{2}\bm{z}^T\bm{P}\bm{z} + \left(\bm{q} + \sum_{p=1}^{n_P} \bm{Q}_p\bm{y}_p \right)^T\bm{z} \\
        \text{s.t.} \; & \textstyle \bm{A}\bm{z} = \bm{b} + \sum_{p=1}^{n_P}\bm{B}_p\bm{y}_p \\
        & \textstyle \bm{G}\bm{z} \leq \bm{h} + \sum_{p=1}^{n_P}\bm{H}_p\bm{y}_p \\
        & \textstyle \bm{z}[1:n_I] \in \{0,1\}^{n_I},
    \end{aligned}
\end{equation}
where $n_P$ and $n_I$ are the numbers of parameters and binary variables, respectively. $\bm{P}$, $\bm{A}$, $\bm{G}$, $\bm{q}$, $\bm{b}$, $\bm{h}$, $\bm{Q}_p$, $\bm{B}_p$, and $\bm{H}_p$ are constants of appropriate dimensions. Note that in \eqref{eq:standard_integer_qp}, the parameters appear on the right-hand sides of both equality and inequality constraints, or as linear coefficients in the objective function. As will be shown in case studies, most mixed-integer convex PSOs, such as $\mathcal{P}_{obf}$, $\mathcal{P}_{obf/sco}$, and $\mathcal{P}_{obf/uncer}$, etc, follow this formulation. Accordingly, \texttt{lapso.optimization} implements an automatic transformation to the standard form \eqref{eq:standard_integer_qp}, upon which the KKT conditions are built as an implicit set of $\bm{z}$ and $\bm{y}_p$s.

In addition, the \texttt{lapso.neuralnet} transforms \texttt{PyTorch}'s \texttt{nn.sequential} modules \cite{paszke2019pytorch} or more complex nested structures composed of piecewise linear layers, e.g. linear, convolutional, max-pooling layers with ReLU activations into mixed-integer linear constraints, e.g. $u(\cdot;\bm{\theta}_u^\star)$ in $\mathcal{P}_{sco}$, $\mathcal{P}_{obf/sco}$ and $\mathcal{M}_{obf/sco}$. In detail, $u({\bm{z},\bm{y};\bm{\theta}_u)} \leq  0$ is equivalent to the following set of constraints,
\begin{equation}\label{eq:linear_mil}
    \begin{aligned}
        \tilde{\bm{z}}_{L} & \leq 0, \\
       \tilde{\bm{z}}_{i+1} & \geq \bm{W}_i\tilde{\bm{z}}_i + \bm{b}_i, \quad \tilde{\bm{z}}_{i+1}  \geq 0, \\
         \bm{u}_i \cdot \bm{v}_i & \geq \tilde{\bm{z}}_{i+1}, \quad \bm{W}_i \tilde{\bm{z}}_i+\bm{b}_i \geq \tilde{\bm{z}}_{i+1}+\left(\bm{1}-\bm{v}_i\right) \circ \bm{l}_i, \\
        \bm{v}_i & \in \{0,1\}^{\left|\bm{v}_i\right|}, \quad \forall i=0,\cdots,L-1,  \\
        \tilde{\bm{z}}_0 & = [\bm{z}^T,\bm{y}^T]^T,
    \end{aligned}
\end{equation}
where $L$ is the number of layers and $(\bm{W}_i,\bm{b}_i)$ are the weights and biases of layer $i$; $\bm{z}_i$ is the output of the $i$-th layer; $\bm{v}_i$ is the auxiliary binary variables; the upper and lower bounds of the $i$-th layer's output is denoted as $\bm{u}_i$ and $\bm{l}_i$, respectively, which are determined by the interval bound propagation (IBP) method \cite{gowal2018effectiveness, xu2023availability}.

Moreover, to benchmark new LAPSO applications, another package \texttt{gridforge} is open-sourced for generating a standard power system testbed from \texttt{PyPower} \cite{zimmerman2011matpower}, enriched with appropriate contextual and power profiles. 
Essentially, for end-to-end machine learning and PSO at the transmission level, spatio-temporally correlated weather, load, and renewable data are needed. After a new grid specification is provided, the raw data is obtained from an open-source dataset for a Texas system \cite{lu2023synthetic} over a year. An automatic data assignment and rescaling process is implemented, subject to distinct generator capacity, transmission line thermal limits, and renewable penetration, etc of the targeted system. 

\section{Case Study Settings, Algorithm Design, and Implementations}\label{sec:case_study}

This section prepares the case-study environment and instantiates the LAPSO primitives (classes of PSO, learning targets, data, hypothesis classes, and training algorithms; cf. Fig.~\ref{fig:learning_pipeline}) for subsequent evaluation under optimization-aware metrics in Section~\ref{sec:simulation}. Unless stated otherwise, all case studies are demonstrated on the IEEE 14-bus system \cite{zimmerman2011matpower}, augmented with four solar units at buses 5, 11, 13, and 14. The network configuration, $\mathcal{P}_{basic}$ formulations, and the preprocessed nodal weather, load, and solar datasets are generated using \texttt{gridforge} for one year at hourly resolution (8{,}760 samples). To focus on algorithmic comparisons, we restrict the main experiments to mixed-integer convex formulations of the form \eqref{eq:standard_integer_qp}. Scalability studies on IEEE 39-, 57-, 118-, and 300-bus systems are reported in Appendix~\ref{app:scale_sco} and Appendix~\ref{app:scale_obf}; extending the demonstrations to other optimization classes (e.g., nonlinear AC OPF) is left for future work.

\subsection{SCO}\label{sec:case_study_sco}

This section describes the simulation setup for the SCO case study in Section~\ref{sec:special_sco}. Using small-signal stability as an example, we define multiple candidate stability assessors and specify how they will be evaluated under the LAPSO pipeline (Fig.~\ref{fig:learning_pipeline}). The LAPSO primitives are defined as follows.

\subsubsection{Class of PSO and Training Target}

A standard mixed-integer linear UC problem, consisting of generator, reserves, and transmission line constraints, is considered as the $\mathcal{P}_{basic}$ in this case study. 
To efficiently assess the grid strength, generalized Short-Circuit Ratio (gSCR) based small-signal stability criterion is considered, e.g.,
\begin{equation}\label{eq:gscr}
    \text{gSCR} := \lambda_{min}\left(\operatorname{diag}\left({\bm{v}_r^2}/{\bm{p}_r}\right)\bm{Y}_{red}\right) \geq \text{gSCR}_{lim},
\end{equation}
in which $\operatorname{diag}\left({\bm{v}_r^2}/{\bm{p}_r}\right)$ is the diagonal matrix of IBR terminal voltage $\bm{v}_r$ and inverter output power $\bm{p}_r$; $\bm{Y}_{red}$ is the reduced nodal admittance matrix after eliminating passive buses and
infinite buses, which is dependent on the topology of the grid, such as generator on/off status $\bm{u}_g$ and the location of IBRs. $\text{gSCR}_{lim}$ is the critical gSCR value, which can be defined by simulation. Detailed derivation can be found in \cite{xu2024incorporation,cui2025control}. 

Although \eqref{eq:gscr} is derived in analytical form, directly encoding into UC with iterative and approximation solution algorithms can cause sub-optimal solutions and non-convergence issues, due to the nonlinearity and nonconvexity of the eigenvalue problem \cite{wang2025synchronous}. Therefore, data-driven approaches are adopted as the stability assessment model. 

\subsubsection{Data}

To formulate the supervised SCO training problem $\mathcal{M}_{sco}$, we construct a labeled dataset of stable and unstable operating points. A sample is labeled as \emph{stable} if \text{gSCR}$\ge \text{gSCR}_{lim}$ and \emph{unstable} otherwise, yielding the binary label $\xi\in\{0,1\}$ used in $\mathcal{M}_{sco}$ where 0 represents the stable label. As shown by \eqref{eq:gscr}, the feature vector includes generator commitment status $\bm{u}_g$ and renewable injections $\bm{p}_r$ (with IBR terminal voltages fixed at $1.0$~p.u. in this study). For the IEEE 14-bus system, we enumerate all non-zero commitment patterns of five generators ($2^5-1$ combinations in total) and discretize each of the four renewable injections into five levels, yielding $(2^5-1)\times 5^4=19{,}375$ labeled samples. For larger systems, we use an active sampling strategy (Appendix~\ref{app:active_sampling}) to concentrate samples near the stability boundary.

\subsubsection{Hypothesis Class and Training Loss}

As a binary classification problem $\mathcal{M}_{sco}$, standard logistic regression (LgR) with binary cross-entropy (BCE) loss is first considered as a candidate stability assessor. Moreover, let $\bm\phi$ denote the feature vector constructed from the operating point (e.g., concatenating $\bm u_g$ and renewable injections); $\mathcal{D}_{stable}$ and $\mathcal{D}_{unstable}$ are the stable and unstable sub-datasets. A constrained logistic regression problem (cLgR) can be solved analytically for $\bm{\theta}_u = [\bm{w}_u,b_u]$,
\begin{equation}\label{eq:cbce}
    \begin{aligned}
        \min_{\bm{w}_u,b_u} & \quad \frac{1}{|\mathcal{D}_{stable}|} \sum_{\bm{\phi}_s\in\mathcal{D}_{stable}} \log \!\left(1 + e^{\bm{w}_u^\top\bm{\phi}_s + b_u}\right) \\
        \text{s.t.} & \quad \bm{w}_u^\top\bm{\phi}_u + b_u \geq 0, \quad \bm{\phi}_u \in \mathcal{D}_{unstable}.
    \end{aligned}
\end{equation}
The objective of \eqref{eq:cbce} is to minimize the entropy loss on the \emph{stable} samples. As a hyperplane is placed between the stable and unstable samples in LgR, a conservative constraint is added in cLgR to ensure that all unstable samples are classified correctly. Although \eqref{eq:cbce} is always feasible, this conservative constraint can increase false positives on stable samples, which may translate into higher operating cost once embedded as $\mathcal{P}_{sco}$; this effect is quantified in Section~\ref{sec:sim_sco}.

In addition, piece-wise linear NN-based stability assessors with different capacities are trained by $\mathcal{M}_{sco}$ as small-signal stability surrogate. 

\subsubsection{Training Algorithm and Metrics}

LgR and cLgR \eqref{eq:cbce} are solved via MP-based method while the NN-based models are solved by SGD. To evaluate the optimization-aware metrics (Fig.~\ref{fig:learning_pipeline}) at the validation stage, the trained stability constraint $u([\bm{u}_g^T, (\bm{p}_r - \bm{p}_{rc})^T]^T;\bm{\theta}_u^\star)\leq 0$ is encoded into $\mathcal{P}_{sco}$ where $\bm{p}_{rc}$ is the renewable curtailment decision. As manually determining the structure of \eqref{eq:linear_mil} and encoding as a set of constraints are time-consuming and prone to error, \texttt{lapso.neuralnet} is used, as shown below.

\begin{python}
# uc: a known unit commitment formulated by CVXPY; 
# NN: trained NN-based stability assessor
from lapso.neuralnet import form_milp
import cvxpy as cp
import numpy as np
# Extract parameters, variables, and constraints
renew = uc.param_dict['renew']
ug, rc = uc.var_dict['ug'], uc.var_dict['rc']
constraints = uc.constraints
# Bounds of the NN input
LB = np.zeros(no_gen + no_solar)
UB = np.concatenate([np.ones(no_gen), renew_max])
IB = (LB[None,:],UB[None,:])
for t in range(T):
    # Return NN as MIL constraints
    nn_constraint, (z,_) = form_milp(NN, IB)
    constraints.extend(nn_constraint)
    # Add stability constraint
    constraints.append(z[-1] <= -1e-3)
    # Link NN input with parameters and variables
    constraints.append(z[0] == cp.hstack([ug[t],
                renew[t] - rc[t]]))
uc_sco = cp.Problem(uc.objective, constraints)
\end{python}

The key usage of \texttt{lapso.neuralnet} is to first extract the relevant parameters and variables ($\bm{p}_r$, $\bm{u_g}$ and $\bm{p}_{rc}$ in this example) of the existing optimization problem $\mathcal{P}_{basic}$ (the UC problem compiled by \texttt{CVXPY}) as the input of the stability assessor. It automatically transforms the trained NN with various types of layers into the MIL form \eqref{eq:linear_mil} with appropriate upper and lower bounds propagated along the neural network. Next, the extracted parameters and decision variables are grouped as input to $\tilde{\bm{z}}_0$ (\texttt{z[0]}) and the NN output $\tilde{\bm{z}}_L$ (\texttt{z[-1]}) is added into new constraints. Therefore, \texttt{lapso.neuralnet} provides a plug-and-play approach to \emph{append} solver-compatible encodings of the trained stability surrogate as additional constraints, enabling modular extension of an existing $\mathcal{P}_{basic}$ model; the resulting optimization-aware evaluation is reported in Section~\ref{sec:sim_sco}.

\subsection{ABF, OBF, and OBF/SCO}\label{sec:obf_case_study}

This section presents the case-study setup for ABF, closed-loop OBF, and their hybrid extension with stability constraints (OBF/SCO). Referring to Fig.~\ref{fig:learning_pipeline}, we first instantiate the LAPSO primitives as follows. 

\subsubsection{Class of PSO and Learning Target}

We consider a continuous dispatch--redispatch (DP--RD) nowcasting setting \cite{vohra2023end} as the downstream PSO chain $\mathcal{P}_{basic}$ for ABF, OBF, and OBF/SCO. Specifically, 
we use the UC formulation with a fixed commitment vector $\bm u_g=[1,1,0,0,1]^\top$,
so that the resulting DP and RD subproblems are convex continuous optimizations. The training target of the forecaster is the predictable renewable trajectory $\bm y_2$ (solar generation), which affects the DP decision and the realized end-to-end operational cost of the DP--RD chain.

\subsubsection{Data, Hypothesis Class, and Training Loss}

\begin{figure}[t]
    \centering
    \includegraphics[width=0.95\linewidth, clip]{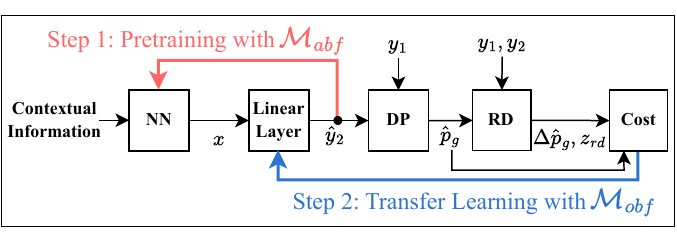}
    \caption{The transfer learning strategy of $\mathcal{M}_{obf}$.}
    \label{fig:obf_exp_framework}
\end{figure}

We use a feedforward NN-based solar forecaster \emph{pretrained} on 8{,}760 hourly samples using the ABF objective $\mathcal{M}_{abf}$ (MSE loss) and SGD. We then apply transfer learning to \emph{fine-tune} the model on weekly data (168 samples) by updating only the final linear layer with parameters $(\bm W_h,\bm b_h)$; cf. Fig.~\ref{fig:obf_exp_framework}. Consequently, the deployed forecaster can be written as a multivariate linear regressor on fixed latent embedding $\bm x$ produced by the frozen part of the pretrained NN, i.e., $\hat{\bm y}_2=\bm W_h\bm x+\bm b_h$. We then form the dataset as $(\bm x^i,\bm y_1^i,\bm y_2^i)\in\mathcal D$, where $\bm y_1^i$ is the exogenous load profile and $\bm y_2^i$ is the realized solar generation. This definition follows Fig.~\ref{fig:obf-uncertain}: $\bm y_1$ is independent of the forecaster, while uncertainty in $\bm y_2$ is associated with the prediction task. 

\subsubsection{Training Algorithm and Metrics}

To fine-tune with $\mathcal{P}_{basic}$, $\mathcal{M}_{obf}$ can be written as \eqref{eq:obf-case-study}, where both DP and RD are represented as the lower-level problems whose objective and constraints are denoted as $f_{dp/rd}(\cdot)$ and $\mathcal{C}_{dp/rd}(\cdot)$, respectively. $\bm{p}_g$ and $\Delta \bm{p}_g$ represent the generation variables in dispatch and redispatch stages; $\bm{z}_{dp}$ and $\bm{z}_{rd}$ denote the remaining variables such as renewable curtailment, etc.
\begin{equation}\label{eq:obf-case-study}
    \begin{aligned}
        \min_{\bm{W}_h,\bm{b}_h} ~ & \frac{1}{|\mathcal{D}|} \sum_{i=1}^{|\mathcal{D}|} f_{pso}(\hat{\bm{p}}_g^{i}, \Delta \hat{\bm{p}}^i_g, \hat{\bm{z}}_{rd}^i)  \\
        \text{s.t.} ~ & \text{For } \forall i=1,\cdots,|\mathcal{D}|, \\
        & \hspace{0.5cm} \Delta \hat{\bm{p}}_g^i, \hat{\bm{z}}_{rd}^i = \arg\min \{ f_{rd}(\Delta\bm{p}_g,\bm{z}_{rd}): \\
        & \hspace{2.2cm} (\Delta\bm{p}_g,\bm{z}_{rd}) \in \mathcal{C}_{rd}(\bm{y}^i_2,\bm{y}^i_1, \hat{\bm{p}}_g^i) \} \\
        & \hspace{0.5cm} \hat{\bm{p}}_g^i, \hat{\bm{z}}_{dp}^i = \arg\min \{ f_{dp}(\bm{p}_g, \bm{z}_{dp}): \\
        & \hspace{2.2cm} (\bm{p}_g, \bm{z}_{dp}) \in \mathcal{C}_{dp}(\hat{\bm{y}}^i_2, \bm{y}_1^i) \} \\
        & \hspace{0.5cm} \hat{\bm{y}}^i_2 = \bm{W}_h\bm{x}^i + \bm{b}_h.
    \end{aligned}
\end{equation}
As illustrated in Fig.~\ref{fig:obf_exp_framework}, the renewable forecast $\hat{\bm{y}}_2^i$ and the given load profile $\bm{y}_1^i$ serve as the input to the DP, whose optimal decision variables are then passed as the inputs to the RD. The upper-level objective integrates the costs of both DP and RD, which becomes the end-to-end cost of the complete chain of PSOs. 
\begin{equation}\label{eq:pso_cost}
    f_{pso}(\hat{\bm{p}}_g^{i}, \Delta \hat{\bm{p}}^i_g, \hat{\bm{z}}_{rd}^i) = \tilde{f}_{dp}(\hat{\bm{p}}_g^{i}) + f_{rd}(\Delta \hat{\bm{p}}^i_g, \hat{\bm{z}}_{rd}^i).
\end{equation}
Apparently, this formulation lies in the \emph{closed-loop} setting discussed in Section~\ref{sec:special_obf}. Moreover, \eqref{eq:obf-case-study} corresponds to the convex $\mathcal{P}_{basic}$ and convex $\bm{h}(:,\bm{\theta}_h)$ case in Table~\ref{tab:obf_train}, which can be solved by MPEC such as with KKT conditions.

The developed \texttt{lapso.optimization.form\_kkt} provides automatic conversion of a given PSO (such as $\mathcal{P}_{basic}$, compiled by \texttt{CVXPY}), to its corresponding KKT formulation. Then the big-M method is applied to linearize the bilinear complementarity slackness conditions into linear mixed-binary conditions. A compact pseudo code for $\mathcal{M}_{obf}$ in \eqref{eq:obf-case-study} using \texttt{lapso} is shown below, where both the lower-level DP and RD problems are reformulated as sets of linearized KKT systems. 

\begin{python}
# (dp,rd): dispatch and redispatch by CVXPY; 
# (X,y): training dataset
from lapso.optimization import form_kkt
import cvxpy as cp
# Define forecaster variables
W = cp.Variable(shape = (X.shape[1], y.shape[1]))
b = cp.Variable(shape = (y.shape[1],))
y_hat = X @ W + b # Renewable forecasting
obj, constraints = 0, []
# For each sample in the dataset
for i in range(X.shape[0]):
    # Convert DP and RD into KKT conditions
    (dp_kkt, dp_var, dp_param)=form_kkt(dp, M=1e4)
    (rd_kkt, rd_var, rd_param)=form_kkt(rd, M=1e4)
    constraints += dp_kkt + rd_kkt
    # Extract parameters in the objective
    P_dp = dp_param['P'][:no_gen, :no_gen]
    q_dp = dp_param['q'][:no_gen]
    P_rd, q_rd = rd_param['P'], rd_param['q']
    # Extract the variables
    z_dp, z_rd = dp_var['x'], rd_var['x']
    # Extract input parameters as linking variables
    dp_link_var = dp_var['param_dict_as_var']
    rd_link_var = rd_var['param_dict_as_var']
    # Link the decision-chain via the constraints
    constraints += [dp_link_var['y'] == y_hat[i],
        rd_link_var['y'] == y[i],
        rd_link_var['pg_parameter'] == z_dp['pg']]
    # Formulate OBF objective
    obj += 0.5*cp.quad_form(z_dp['pg'], P_dp) 
        + q_dp@z_dp['pg'] + 0.5*cp.quad_form(
            cp.hstack(list(z_rd),P_rd) 
        + q_rd@cp.hstack(list(z_rd))
prob = cp.Problem(cp.Minimize(obj_all), constraints)
\end{python}

In implementation, \texttt{lapso.optimization.form\_kkt} returns (i) a set of KKT constraints (with big-$M$ linearization of complementarity) and (ii) structured dictionaries for primal/dual variables and parameters; the OBF training problem is built by appending these KKT constraints and adding linking constraints that connect the forecaster output to DP parameters and DP decisions to RD parameters. 
As a practical PSO is formulated in a much more complex manner than its standard form \eqref{eq:standard_integer_qp}, by using \texttt{form\_kkt}, tedious manual structural identification, KKT formulation, and parameter extraction can be avoided.

When $\mathcal{M}_{obf/sco}$ is considered, a learned stability constraint (e.g., from LgR as a linear inequality) is appended to both DP and RD subproblems, yielding a stability-constrained PSO chain. The construction then follows the same procedure as the OBF case above.

\subsection{OBF/Uncer}\label{sec:obf_uncer_case_study}

In this section, the uncertainty quantification framework in Section~\ref{sec:obf-uncertain} is instantiated and countermeasure on Opt-Uncertainty $\mathcal{M}_{obf/uncer}$ is formulated as robust problem. A worst-case analysis for both ML-Uncertainty and Opt-Uncertainty is formulated in Appendix~\ref{app:multi_uncertainty}.

In \eqref{eq:obf-case-study}, the loads are assumed to be exactly known during the $\mathcal{M}_{obf}$ training so that $\bm{y}^i_1$s are the same for both DP and RD. In the setting of $\mathcal{M}_{obf/uncer}$, at the RD stage, it is assumed that the load is subject to a bounded sample-dependent $\ell_{\infty}$ uncertainty set, e.g.,  $\bm{y}_{1,rd}^i\in \mathcal{Y}^i =  [\underline{\bm{y}}_{1,rd}^i, \bar{\bm{y}}_{1,rd}^i],\forall i=1,\cdots,|\mathcal{D}|$. Note that when $(\bm{W}_h,\bm{b}_h)$ are fixed, \eqref{eq:obf-case-study} becomes single level RD as $\hat{\bm{p}}_g^i,\forall i=1,\cdots,|\mathcal{D}|$ in the DP can be optimized explicitly for given renewable forecast $\hat{\bm{y}}^i_2 = \bm{W}_h\bm{x}^i + \bm{b}_h$. Therefore, the worst operational cost for each sample $i$ subject to uncertainty set $\mathcal{Y}^i$ is obtained independently as the following maximization problem,
\begin{equation}\label{eq:obf-case-worst}
    \begin{aligned}
        \mathcal{Q}(\hat{\bm{p}}_g^i) := & \underbrace{\tilde{f}_{dp}(\hat{\bm{p}}_g^i)}_{\text{Const.}} + \max_{\bm{y}_{1,rd}^i\in\mathcal{Y}^i} f_{rd}(\Delta \hat{\bm{p}}_g^i, \hat{\bm{z}}_{rd}^i) \\
        \text{s.t.} & \; \Delta \hat{\bm{p}}_g^i, \hat{\bm{z}}_{rd}^i = \arg\min \{ f_{rd}(\Delta\bm{p}_g,\bm{z}_{rd}): \\
        & \hspace{1.7cm} (\Delta\bm{p}_g,\bm{z}_{rd}) \in \mathcal{C}_{rd}(\bm{y}^i_2, \bm{y}^i_{1,rd}, \hat{\bm{p}}_g^i) \},
    \end{aligned}
\end{equation}
with given $\hat{\bm{p}}_g^i,\forall i=1,\cdots,|\mathcal{D}|$ and $\tilde{f}_{dp}(\hat{\bm{p}}_g^i)$ represents the contribution of $\hat{\bm{p}}_g^i$ to the PSO cost in \eqref{eq:pso_cost}. 

Considering a ``wait-and-see'' setting to hedge against the worst-case load realization at RD \eqref{eq:obf-case-worst}, one formulation of $\mathcal{M}_{obf/uncer}$ can be designed as,
\begin{equation}
\label{eq:obf-case-study-robust}
    \begin{aligned}
        \min_{\bm{W}_h,\bm{b}_h} \; & \frac{1}{|\mathcal{D}|} \sum_{i=1}^{|\mathcal{D}|} \tilde{f}_{dp}(\hat{\bm{p}}_g^{i}) + \max_{\bm{y}_{1,rd}^i\in\mathcal{Y}^i} f_{rd}(\Delta \hat{\bm{p}}_g^i, \hat{\bm{z}}_{rd}^i)  \\
        \text{s.t.} \; & \text{For } \forall i=1,\cdots,|\mathcal{D}|, \\
        & \hspace{0.3cm} \Delta \hat{\bm{p}}_g^i, \hat{\bm{z}}_{rd}^i = \arg\min \{ f_{rd}(\Delta\bm{p}_g,\bm{z}_{rd}): \\
        & \hspace{2.2cm} (\Delta\bm{p}_g,\bm{z}_{rd}) \in \mathcal{C}_{rd}(\bm{y}^i_2, \bm{y}^i_{1,rd}, \hat{\bm{p}}_g^i) \} \\
        & \hspace{0.3cm} \hat{\bm{p}}_g^i, \hat{\bm{z}}_{dp}^i = \arg\min \{ f_{dp}(\bm{p}_g, \bm{z}_{dp}): \\
        & \hspace{2.2cm}  (\bm{p}_g, \bm{z}_{dp}) \in \mathcal{C}_{dp}(\hat{\bm{y}}^i_2, \bm{y}^i_1) \} \\
        & \hspace{0.3cm} \hat{\bm{y}}^i_2 = \bm{W}_h\bm{x}^i + \bm{b}_h.
    \end{aligned}
\end{equation}
which turns into a two-stage bilevel robust optimization. Similarly to $\mathcal{M}_{obf}$, the lower level DP is first represented by the KKT conditions (denoted as $\mathcal{KKT}_{dp}(\bm{y}_1^i,\hat{\bm{y}}_2^i)$) and linearized by the big-M method so that \eqref{eq:obf-case-study-robust} becomes,
\begin{equation}
\label{eq:obf-case-study-robust_1}
    \begin{aligned}
        \min_{\bm{W}_h,\bm{b}_h} \; & \frac{1}{|\mathcal{D}|} \sum_{i=1}^{|\mathcal{D}|} \{\tilde{f}_{dp}(\hat{\bm{p}}_g^{i})  \\
        & \hspace{0.2cm} + \max_{\bm{y}_{1,rd}^i\in\mathcal{Y}^i} \min_{\Delta \bm{p}_g^i, \bm{z}_{rd}^i \in \mathcal{C}_{rd}(\bm{y}_2^i, \bm{y}_{1,rd}^i, \hat{\bm{p}}_g^i)} f_{rd}(\Delta \bm{p}_g^i, \bm{z}_{rd}^i) \}  \\
        \text{s.t.} \; & \text{For } \forall i=1,\cdots,|\mathcal{D}|, \\
        & \hspace{0.3cm} \hat{\bm{p}}_g^i, \hat{\bm{z}}_{dp}^i \in \mathcal{KKT}_{dp}(\bm{y}_1^i, \hat{\bm{y}}_2^i), ~ \hat{\bm{y}}^i_2 = \bm{W}_h\bm{x}^i + \bm{b}_h.
    \end{aligned}
\end{equation}

This paper solves \eqref{eq:obf-case-study-robust_1} using a column-and-constraint generation (C\&CG) procedure adapted from \cite{zeng2013solving}. In the worst case, the number of iterations can grow with the number of extreme points of the uncertainty set $\mathcal{Y}^i$; in practice, convergence is typically achieved in far fewer iterations. For clarity, the overall workflow is summarized in Algorithm~\ref{alg:ccg_obf_uncer}, while the detailed master problem and subproblem formulations are given in \eqref{eq:obf-case-study-main} and \eqref{eq:obf-case-worst_1}, respectively. 

In Algorithm~\ref{alg:ccg_obf_uncer}, the subproblems in \eqref{eq:obf-case-worst_1} are independent across samples and can therefore be solved in parallel. If the gap between the updated upper and lower bounds exceeds the tolerance $\epsilon$, new optimality cuts associated with the worst-case load realizations are added to the master problem. 
In implementation, the KKT systems in both the master problem and the subproblems are generated automatically using \texttt{lapso.optimization.form\_kkt} function.

\begin{equation}\label{eq:obf-case-study-main}
    \begin{aligned}
        \min_{\bm{W}_h, \bm{b}_h, \eta^i, \hat{\bm{p}}_g^i,\forall i} \; & \frac{1}{|\mathcal{D}|} \sum_{i=1}^{|\mathcal{D}|} \tilde{f}_{dp}(\hat{\bm{p}}_g^{i}) + \eta^i  \\
        \text{s.t.} 
        \; & \text{For } \forall l=1,\cdots, k, \forall i=1,\cdots,|\mathcal{D}|,  \\
        & \hspace{0.3cm} \eta^i \geq f_{rd}(\Delta\bm{p}_g^{i,(l)}, \bm{z}_{rd}^{i,(l)}), \; \eta^i \geq 0\\
        & \hspace{0.3cm} (\Delta\bm{p}_g^{i,(l)}, \bm{z}_{rd}^{i,(l)}) \in \mathcal{C}_{rd}(\bm{y}^i_2,\bm{y}_{1,rd}^{i,(l),\star}, \hat{\bm{p}}_g^i) \\
        & \hspace{0.3cm} \hat{\bm{p}}_g^i, \hat{\bm{z}}_{dp}^i \in \mathcal{KKT}_{dp}(\bm{y}_1^i, \hat{\bm{y}}_2^i)
        \\
        & \hspace{0.3cm} \hat{\bm{y}}^i_2 = \bm{W}_h\bm{x}^i + \bm{b}_h,
    \end{aligned}
\end{equation}

\begin{equation}\label{eq:obf-case-worst_1}
            \begin{aligned}
                \mathcal{Q}(\hat{\bm{p}}_g^{i,(k+1),\star}) & := \underbrace{\tilde{f}_{dp}(\hat{\bm{p}}_g^{i,(k+1),\star})}_{\text{Const.}} + \max_{\bm{y}_{1,rd}^i\in\mathcal{Y}^i} f_{rd}(\Delta \hat{\bm{p}}_g^i, \hat{\bm{z}}_{rd}^i) \\
                \text{s.t.} & \; (\Delta \hat{\bm{p}}^i_g, \hat{\bm{z}}^i_{rd}) \in \mathcal{KKT}_{rd}(\bm{y}^i_2, \bm{y}^i_{1,rd}, \hat{\bm{p}}_g^{i,(k+1),\star}).
            \end{aligned}
\end{equation}

\begin{algorithm}[t]
\small
\caption{C\&CG for solving $\mathcal{M}_{obf/uncer}$}
\label{alg:ccg_obf_uncer}
\SetKwInOut{Input}{Input}
\SetKwInOut{Output}{Output}
\Input{Training dataset $\mathcal{D}$, uncertainty sets $\{\mathcal{Y}^i\}_{i=1}^{|\mathcal{D}|}$, tolerance $\epsilon$.}
\Output{Robust forecaster parameters $(\bm{W}_h^\star,\bm{b}_h^\star)$}

Initialize $LB=-\infty$, $UB=+\infty$, and iteration counter $k=0$.

\Repeat{$UB-LB\le\epsilon$}{
    Solve the master problem \eqref{eq:obf-case-study-main} and obtain
    $(\bm{W}_h^{(k+1),\star}, \bm{b}_h^{(k+1),\star}, \eta^{(k+1),\star}, \hat{\bm{p}}_g^{i,(k+1),\star}, \forall i)$\;
    
    Update the lower bound:
    \[
    LB=\textstyle\frac{1}{|\mathcal{D}|}\sum_{i=1}^{|\mathcal{D}|}
    \left(\tilde{f}_{dp}(\hat{\bm{p}}_g^{i,(k+1),\star})+\eta^{i,(k+1),\star}\right)
    \]
    
    \For{$i=1,\dots,|\mathcal{D}|$}{
        Solve the subproblem \eqref{eq:obf-case-worst_1} for given $\hat{\bm{p}}_g^{i,(k+1),\star}$ and obtain worst-case realization $\bm{y}_{1,rd}^{i,(k+1),\star}$ together with the corresponding value $\mathcal{Q}(\hat{\bm{p}}_g^{i,(k+1),\star})$\;
    }
    
    Update the upper bound:
    \[
    UB=\min\!\left\{
    UB,\;
    \textstyle\frac{1}{|\mathcal{D}|}\sum_{i=1}^{|\mathcal{D}|}
    \mathcal{Q}(\hat{\bm{p}}_g^{i,(k+1),\star})
    \right\}
    \]
    
    \If{$UB-LB>\epsilon$}{
        \For{$i=1,\dots,|\mathcal{D}|$}{
            Add the optimality cut associated with $\bm{y}_{1,rd}^{i,(k+1),\star}$ to the master problem:
            \[
            \eta^i \ge f_{rd}(\Delta\bm{p}_g^{i,(k+1)}, \bm{z}_{rd}^{i,(k+1)})
            \]
            \[
            (\Delta\bm{p}_g^{i,(k+1)}, \bm{z}_{rd}^{i,(k+1)})
            \in
            \mathcal{C}_{rd}(\bm{y}_2^i,\bm{y}_{1,rd}^{i,(k+1),\star},\hat{\bm{p}}_g^i)
            \]
        }
        Set $k \leftarrow k+1$\;
    }
}
Return $(\bm{W}_h^{(k+1),\star},\bm{b}_h^{(k+1),\star})$\;
\end{algorithm}

\section{Simulation Results}\label{sec:simulation}

This section presents the numerical results by following LAPSO pipeline (Fig.~\ref{fig:learning_pipeline}), based on the definition of primitives and algorithms in Section~\ref{sec:case_study}. All optimizations are run on two Intel\textregistered Xeon\textregistered Gold 6230R CPUs (2.10\,GHz), and NNs are trained on an NVIDIA GeForce RTX 3090 GPU.

\subsection{Small-Signal Stability Constrained Optimization}\label{sec:sim_sco}

\begin{table*}[t]
    \centering
    \footnotesize
    \caption{Performance of SCO with different stability assessors, averaged over 365 days. \texttt{MOSEK} is used as the solver backend.}
    \begin{tabular}{c|c|c|c|c|c|c|c|c|c}
         \multicolumn{2}{c|}{ \textbf{Type}} & \textbf{Ave. Cost} ($\pounds$) & \textbf{UR} (\%) & \textbf{SR} (\%) & \textbf{DR} (\%) & \textbf{OR} (\%) & \textbf{No. Param.} & \textbf{No. Binary} & \textbf{Ave. Time} ($s$)   \\\hline
         \multicolumn{2}{c|}{ $\mathcal{P}_{basic}$ } & 15984.50 & 96.97 & \multicolumn{4}{c|}{N/A} & $15\times24 = 360$ & 0.467 \\\hline
       \multirow{6}{*}{$\mathcal{P}_{sco}$} & \textbf{LgR} & 17941.25 & 6.61 & 98.00 & 0.00 & 2.85 & 10 & $(15+0)\times24 = 360$ & 1.151 \\\cline{2-10}
        & \textbf{cLgR} & 20391.25 & 0.00 & 100.00 & 0.00 & 12.81 & 10 & $(15+0)\times24 = 360$ & 0.896  \\\cline{2-10}
        & $\bm{NN}_2^{12}$ & 17944.49 & 6.34 & 98.13 & 0.00 & 2.85 & 12 & $(15+1)\times24 = 384$ & 1.165 \\\cline{2-10}
        &  $\bm{NN}_2^{111}$ & 17628.69 & 6.34 & 98.58 & 0.00 & 0.68 & 111 & $(15+10)\times24 = 600$ & 5.980 \\\cline{2-10}
        & $\bm{NN}_3^{161}$ & 17645.20 & 0.28 & 99.95 & 0.00 & 1.03 & 161 & $(15+15)\times24 = 720$ & 15.598 \\\cline{2-10}
        & $\bm{NN}_3^{221}$ & 17667.87 & 0.00 & 100.00 & 0.00 & 0.52 & 221 & $(15+20)\times24 = 840$ & 54.752 
    \end{tabular}
    \label{tab:sco}
\end{table*}

To start, the models considered in this section include
\begin{itemize}[leftmargin=*]
    \item \textbf{LgR}: Logistic regression with standard BCE loss.
    \item \textbf{cLgR}: Logistic regression with constrained BCE loss \eqref{eq:cbce}.
    \item $\textbf{NN}_{\bm{L}}^{\bm{\psi}}$: NN with $\bm{L}$ number of linear layers and $\bm{\psi}$ number of trainable parameters.
\end{itemize}

Following LAPSO pipeline (Fig.~\ref{fig:learning_pipeline}), for each candidate assessor, we (i) train the model on $\mathcal{M}_{sco}$; (ii) append its solver-compatible encoding to $\mathcal{P}_{basic}$ to form $\mathcal{P}_{sco}$; (iii) solve $\mathcal{P}_{sco}$ across the evaluation horizon; and (iv) compute specialized optimization-aware metrics (defined below). Model selection is then performed in a validation-driven manner by enforcing operational acceptability criteria and selecting the best-performing candidate among the acceptable set (Table~\ref{tab:lapso_coupling_guidance}).

The $\mathcal{P}_{basic}$ in this case study optimizes over 24 hours, resulting in 360 binary variables. When the NN-based stability constraint is encoded as \eqref{eq:linear_mil}, each ReLU unit introduces one additional binary variable per time period; hence the total number of binary variables increases proportionally with the number of ReLU units and the horizon length. $\mathcal{P}_{sco}$ is solved by \texttt{MOSEK} with absolute and relative optimality tolerance gaps set to 0.001. 

\begin{figure}[h]
    \centering
    \includegraphics[width=0.60\linewidth]{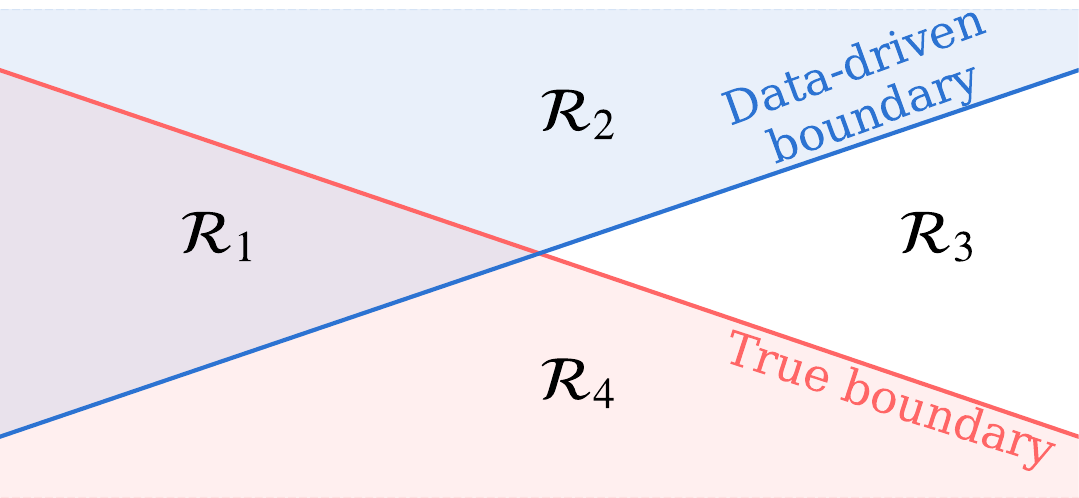}
    \caption{Illustrative areas for stable and unstable operational points. 
    The red and blue areas are the stable regions for data-driven gSCR assessor and true gSCR criterion \eqref{eq:gscr}.}
    \label{fig:sco_area}
\end{figure}

To start, four regions ($\mathcal{R}_1$-$\mathcal{R}_4$) are defined in Fig.~\ref{fig:sco_area}. For the given load and renewable profiles, two sets of operation points can be obtained by solving $\mathcal{P}_{basic}$ and $\mathcal{P}_{sco}$, whose index sets are denoted as $\mathcal{D}^{basic}$ and $\mathcal{D}^{sco}$, respectively. $\mathcal{D}^{basic}$ and $\mathcal{D}^{sco}$ are further assigned to the different regions, based on their stability performances by the learned data-driven and true gSCR stability indexes.

Note that $\mathcal{D}^{basic}$ can belong to any of the four regions while operation points $\mathcal{D}^{sco}$ can only belong to $\mathcal{R}_1$ and $\mathcal{R}_2$ due to the existence of the stability constraint. We further denote $\mathcal{R}_{1,2,3,4}^{basic}$ and $\mathcal{R}_{1,2}^{sco}$ as the index set of operation points assigned to different regions for $\mathcal{P}_{basic}$ and $\mathcal{P}_{sco}$, respectively. The following optimization-aware metrics, such as Unstable Rate (UR), Stabilization Rate (SR), Destabilization Rate (DR), and Overreaction Rate (OR) are defined in \eqref{eq:sco_metric}. 

In detail, UR and OR measure \emph{static} properties of the induced operating points. Specifically, UR evaluates the percentage of unstable outcomes after solving the operational problem, while OR quantifies conservatism by measuring the fraction of truly stable operating points excluded by the learned stability boundary.
\begin{equation}\label{eq:sco_metric}
    \begin{aligned}
        \text{UR}^{basic} & = |\mathcal{R}_2^{basic} \cup \mathcal{R}_3^{basic} | / |\mathcal{D}^{basic}|\\
        \text{UR}^{sco} & = |\mathcal{R}_2^{sco} | / |\mathcal{D}^{sco}|\\
        \text{SR} & = |(\mathcal{R}_2^{basic} \cup \mathcal{R}_3^{basic}) \cap\mathcal{R}_1^{sco} | / |\mathcal{R}_2^{basic} \cup \mathcal{R}_3^{basic}| \\
        \text{DR} & = |(\mathcal{R}_1^{basic} \cup \mathcal{R}_4^{basic}) \cap\mathcal{R}_2^{sco} | / |\mathcal{R}_1^{basic} \cup \mathcal{R}_4^{basic}| \\
        \text{OR} & = |\mathcal{R}_4^{basic}| / |\mathcal{R}_1^{basic} \cup \mathcal{R}_4^{basic}|
    \end{aligned}
\end{equation}

 In contrast, SR and DR measure \emph{transition} properties when moving from $\mathcal{P}_{basic}$ to $\mathcal{P}_{sco}$: SR captures how often originally unstable operating points are corrected by the added stability constraint, whereas DR captures how often originally stable operating points become unstable after re-optimization. In this case study, UR is evaluated on a daily basis, while the remaining metrics are evaluated hourly. These metrics are deliberately defined at the level of operational outcomes rather than standalone classification outputs. Conventional predictive metrics such as true positive rate (TPR) and false positive rate (FPR) only assess the assessor in isolation and cannot capture how the learned boundary interacts with the downstream optimization, which is exactly the quantity of interest under the LAPSO view.

The average performance metrics are summarized in Table~\ref{tab:sco}. Referring to the optimization-aware metrics defined in Fig.~\ref{fig:learning_pipeline}, \emph{computational tractability} is assessed by the Averaged Solve Time; (ii) \emph{constraint satisfaction} is assessed by UR, SR, and DR; (iii) \emph{economic performance} is assessed by the realized operational cost, with OR serving as a conservatism indicator that helps interpret cost increases; and (iv) \emph{solution quality indicators} are assessed via solver status and, for mixed-integer formulations, the reported optimality gap. In this case study, optimal statuses are returned within predefined convergence tolerance gaps in \texttt{MOSEK} for all samples.

\begin{figure}[t]
    \centering
     \begin{subfigure}[b]{0.49\linewidth}
         \centering
         \includegraphics[width=\textwidth]{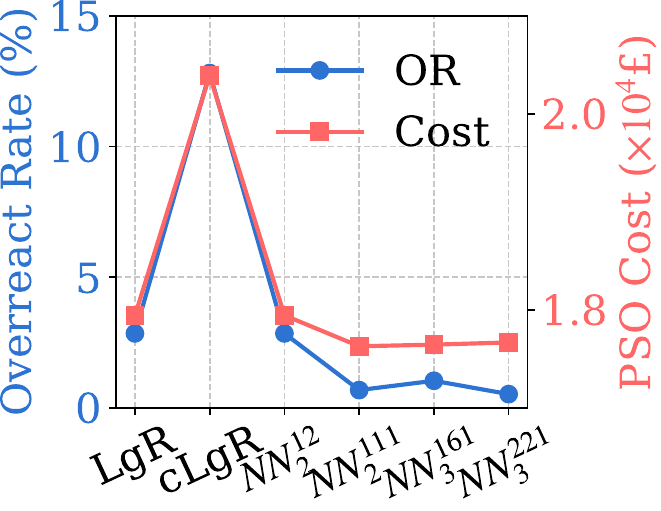}
     \end{subfigure}
     \hfill
     \begin{subfigure}[b]{0.49\linewidth}
         \centering
         \includegraphics[width=\textwidth]{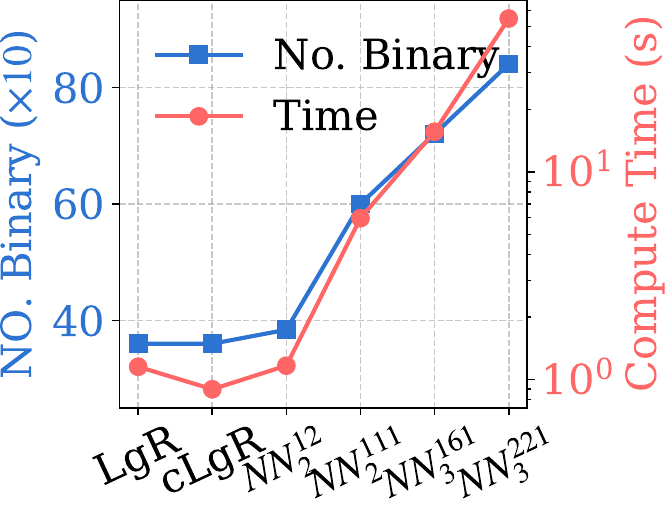}
     \end{subfigure}
        \caption{Performances on different $\mathcal{M}_{sco}$ models.}
        \label{fig:sco_compare}
\end{figure}

The results indicate that all $\mathcal{P}_{sco}$ models reduce the unstable rate to below 7\%, but they do so through different trade-offs among stability, cost, and computation time. Even the simple LgR assessor stabilizes about 98\% of the originally unstable samples in $\mathcal{P}_{basic}$, showing that a low-complexity learned constraint can already provide substantial security improvement. However, eliminating all unstable samples is not cost-free: the constrained cLgR model, despite having the same structural complexity as LgR, introduces an OR of 12.81\%, which leads to a marked increase in operating cost. This confirms the SCO design intuition in Fig.~\ref{fig:sco_design}: a more conservative assessor can improve security, but may do so by excluding too many truly safe operating points. Compared with the linear models, the NN-based assessors provide a better security--economy trade-off. As model expressiveness increases, UR decreases and SR improves, while the increase in cost remains moderate for the better-performing candidates. The price paid is computational: more expressive NNs introduce more binary variables after encoding, which increases solve time. Fig.~\ref{fig:sco_compare} makes both trends explicit by showing a strong positive correlation between OR and operational cost, and between the number of binary variables and computation time.

From the LAPSO model-selection perspective, these results show why predictive performance alone is insufficient. The preferred ML assessor is not the one that is merely most conservative or most expressive, but the one that best balances \emph{constraint satisfaction}, \emph{economic performance}, and \emph{computational tractability}. Under these criteria, $NN_{3}^{161}$ emerges as a favorable choice among the candidates considered: it achieves very low UR and high SR while maintaining acceptable solve time and competitive cost, as by Table~\ref{tab:lapso_coupling_guidance}.

\subsection{Objective-based Renewable Forecasting}\label{sec:sim_obf}

Simulation results based on the settings in Section~\ref{sec:obf_case_study} are reported in this section. We evaluate how different training problems (ABF: $\mathcal{M}_{abf}$; closed-loop OBF: $\mathcal{M}_{obf}$; and the robust variant: $\mathcal{M}_{obf/uncer}$) perform under two deployment settings, namely $\mathcal{P}_{obf}$ and its stability-constrained counterpart $\mathcal{P}_{obf/sco}$. 

\begin{figure}[h]
    \centering
    \includegraphics[width=0.84\linewidth]{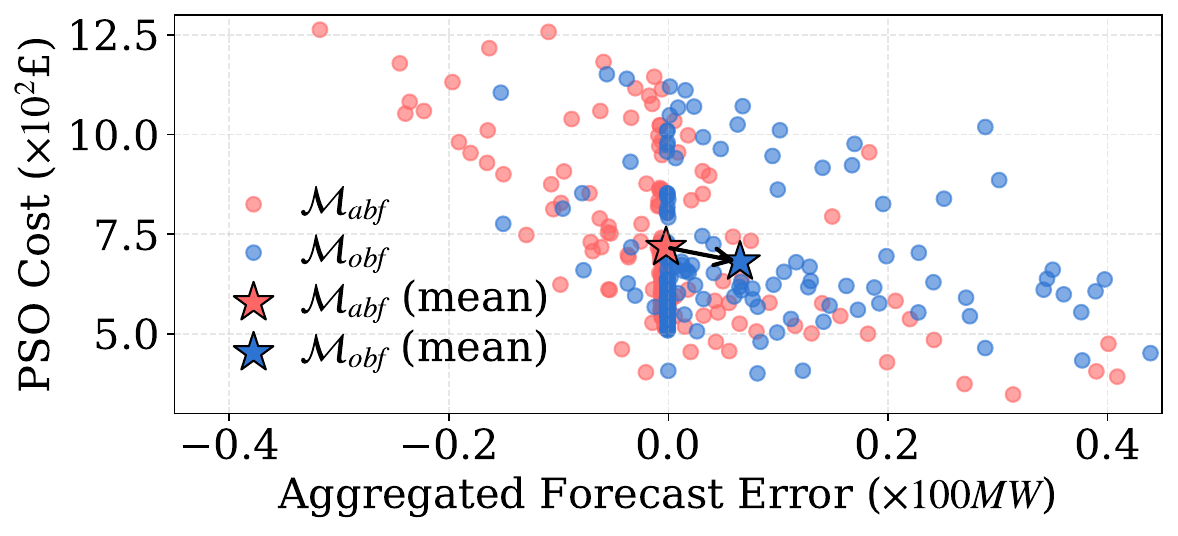}
    \caption{Scatter plots for PSO cost and the aggregated forecast error (computed as aggregated true solar minus solar forecast) under $\mathcal{M}_{obf}$ and $\mathcal{M}_{abf}$. Each point represents one sample in week 30.}
    \label{fig:obf_scatter}
\end{figure}

\begin{table*}[t]
    \centering
    \footnotesize
    \caption{Averaged performance of ABF, OBF, and OBF/SCO over 8760-hour load-renewable profiles.}
    \begin{tabular}{c|c|c|c|c|c|c|c|}
         \multirow{2}{*}{$\mathcal{M}$} &  \multirow{2}{*}{\textbf{MAPE} ($\%$)}   & \multicolumn{3}{c|}{$\mathcal{P}_{obf}$} &   \multicolumn{3}{c|}{$\mathcal{P}_{obf/sco}$}\\\cline{3-8}
         &  & \textbf{Ave. Cost} ($\pounds$) & \textbf{DP-UR} (\%) &  \textbf{RD-UR} (\%)  & \textbf{Ave. Cost} ($\pounds$) & \textbf{DP-UR} (\%) &  \textbf{RD-UR} (\%) \\\hline
        True & 0.00 & 456.36 & 6.42 & 6.85 & 590.35 & 0.00 & 0.00 \\
        $\mathcal{M}_{abf}$ & 6.89 & 491.52 & 6.05 & 6.85 & 621.83 & 0.00 & 0.00 \\
        $\mathcal{M}_{obf}$ & 22.68 & 474.21 & 6.10 &  6.85 & 608.45 & 0.00 & 0.00 \\
        $\mathcal{M}_{obf/sco}$ & 19.51 & 475.74 & 5.21 & 6.85 & 605.34 & 0.00 & 0.00
    \end{tabular}
    \label{tab:obf_result}
\end{table*}

Fig.~\ref{fig:obf_scatter} illustrates the per-hour operational cost against the corresponding renewable forecaster error. The forecaster trained by $\mathcal{M}_{obf}$ tends to underestimate renewable generation so as to reduce expensive real-time redispatch and reserve corrections. This behavior makes clear why predictive accuracy and system-level cost need not align and a higher MAPE does not contradict a lower realized operational cost.

The hourly performances of $\mathcal{P}_{obf}$ and $\mathcal{P}_{obf/sco}$ over one year are reported in Table~\ref{tab:obf_result}. Using the true solar profile provides a natural lower-bound benchmark for realized operational cost under the same deployment structure. Overall, all forecasters can stabilize the grid after encoding the stability constraint for both DP and RD (i.e., by transforming from $\mathcal{P}_{obf}$ to $\mathcal{P}_{obf/sco}$); however, forecasters trained by different methods inevitably increase the operational costs by different amounts. As expected, the forecaster trained by $\mathcal{M}_{abf}$ achieves the best predictive accuracy. The forecaster trained by $\mathcal{M}_{obf}$ can reduce the PSO cost by 3.5\% when evaluated on $\mathcal{P}_{obf}$ and by 2.0\% on $\mathcal{P}_{obf/sco}$, compared to $\mathcal{M}_{abf}$. When $\mathcal{M}_{obf/sco}$ is implemented, the reduction in $\mathcal{P}_{obf/sco}$ cost increases to 2.6\% and the cost increase in $\mathcal{P}_{obf}$ against $\mathcal{M}_{obf}$ is limited. These results demonstrate a degree of transferability between the forecaster trained by $\mathcal{M}_{obf}$ and $\mathcal{M}_{obf/sco}$. The results also show that the training objectives of $\mathcal{M}_{obf/sco}$ and $\mathcal{M}_{obf}$ are more aligned compared to those of $\mathcal{M}_{obf}$ and $\mathcal{M}_{abf}$.

\begin{figure}
     \centering
     \begin{subfigure}[b]{0.45\linewidth}
         \centering
         \includegraphics[width=\textwidth]{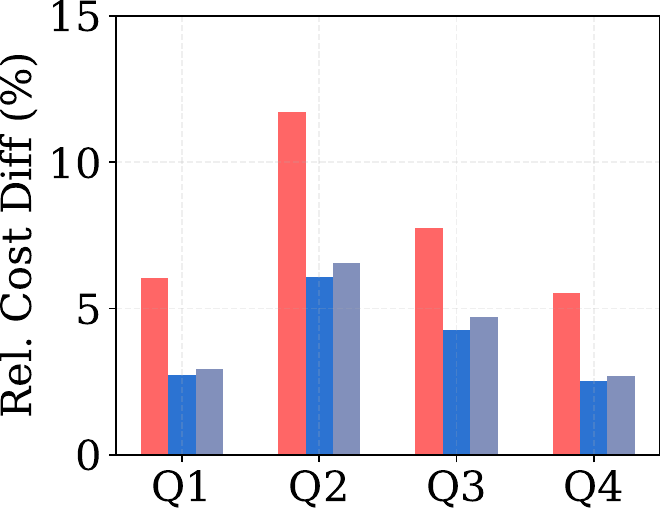}
         \caption{$\mathcal{P}_{obf}$.}
     \end{subfigure}
     \hspace{0.5cm}
     \begin{subfigure}[b]{0.45\linewidth}
         \centering
         \includegraphics[width=\textwidth]{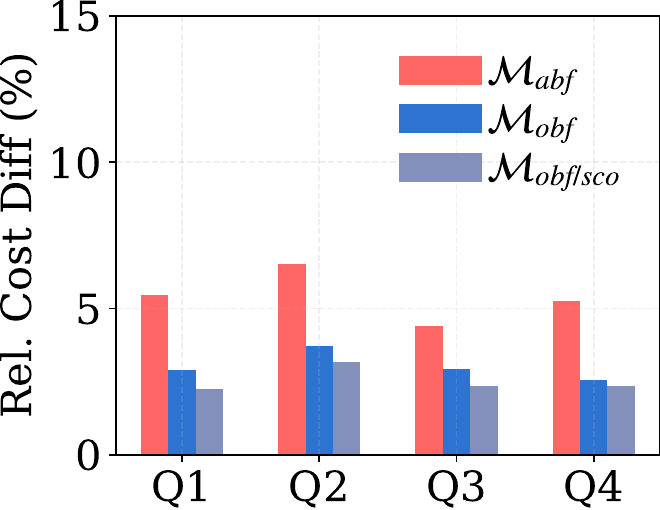}
         \caption{$\mathcal{P}_{obf/sco}$.}
     \end{subfigure}
        \caption{Seasonal performances on different OBF training methods. The relative cost difference is measured against the operational cost driven by true renewable generations. Q1-Q4 represent the quarters in the selected year.}
        \label{fig:obf_season}
\end{figure}

\begin{figure}
    \centering
    \includegraphics[width=0.80\linewidth]{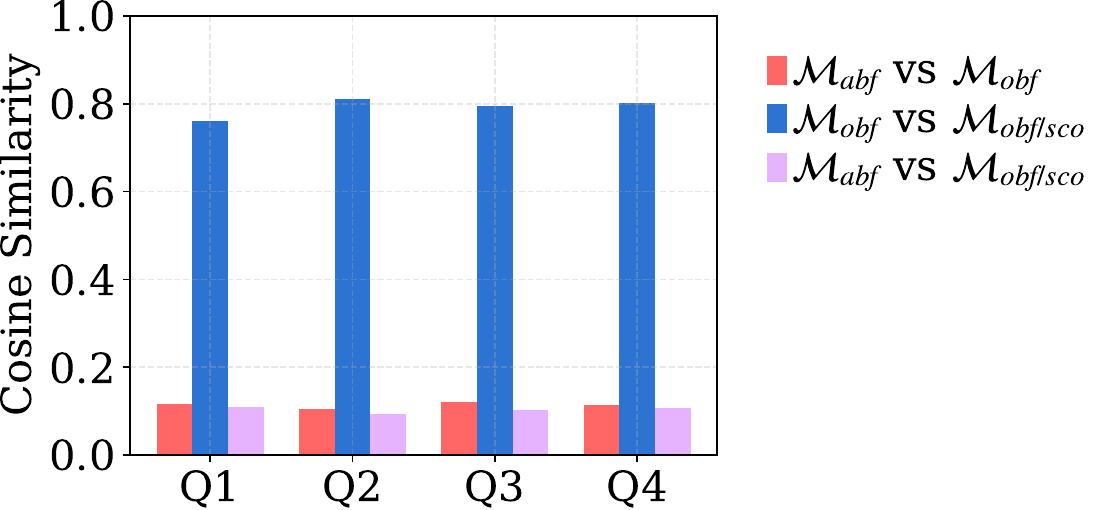}
    \caption{Averaged sample-wise cosine similarities between different training methods.}
    \label{fig:cosine}
\end{figure}

The relative cost differences of different training methods with respect to the cost on true renewable generation are shown in Fig.~\ref{fig:obf_season}. In both $\mathcal{P}_{obf}$ and $\mathcal{P}_{obf/sco}$, the cost differences are larger during the summer terms, which also suggests that the OBF is more effective when renewable penetration is high. Moreover, the cost difference in $\mathcal{P}_{obf/sco}$ is smaller than $\mathcal{P}_{obf}$ due to the existence of stability constraint. To assess cross-setting alignment, Fig.~\ref{fig:cosine} reports the averaged sample-wise cosine similarity between the gradients in \eqref{eq:grad} induced by different training objectives (Section~\ref{sec:obf-uncertain}). The similarity between $\mathcal{M}_{obf/sco}$ and $\mathcal{M}_{obf}$ is confirmed to be much higher than that of the others, implying that the training objectives are more likely to be reduced by one another. 

\begin{figure*}
     \centering
     \begin{subfigure}[b]{0.32\linewidth}
         \centering
         \includegraphics[width=\textwidth]{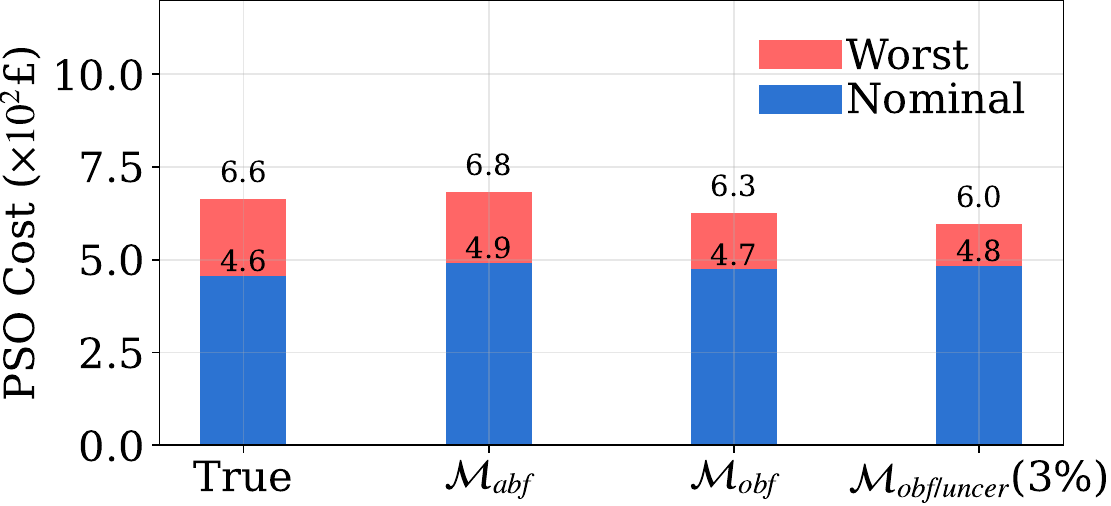}
         \caption{3\% Load Variation}
     \end{subfigure}
     \hfill
     \begin{subfigure}[b]{0.32\linewidth}
         \centering
         \includegraphics[width=\textwidth]{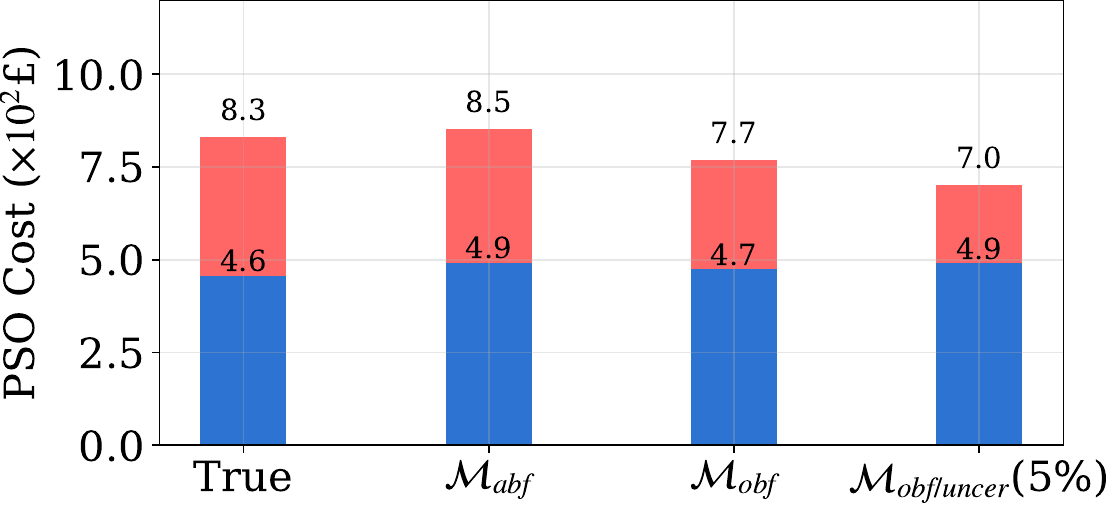}
         \caption{5\% Load Variation}
     \end{subfigure}
     \hfill
     \begin{subfigure}[b]{0.32\linewidth}
         \centering
         \includegraphics[width=\textwidth]{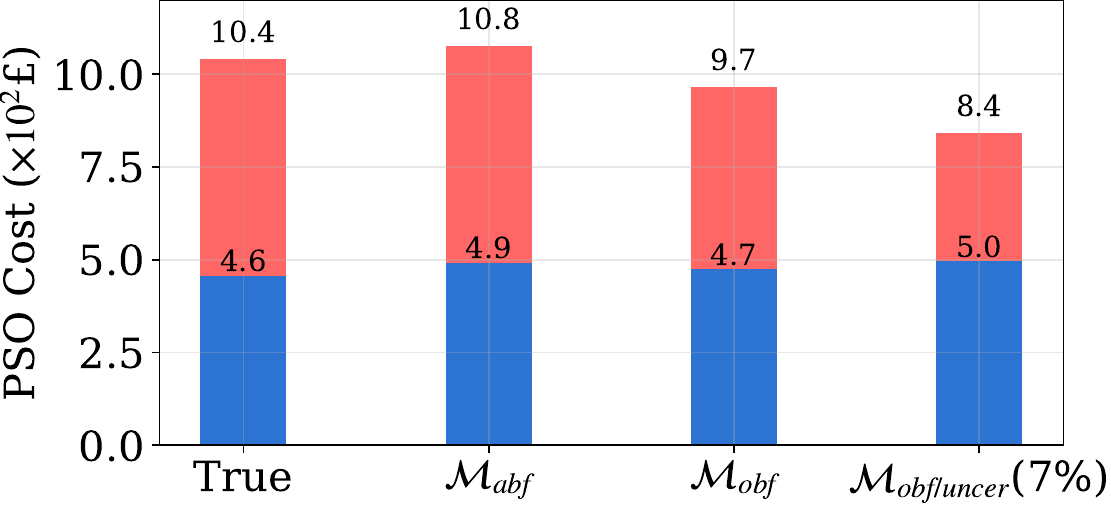}
         \caption{7\% Load Variation}
     \end{subfigure}
        \caption{Performances of different OBF training algorithms under various load variation levels. $\mathcal{M}_{obf/uncer}(\alpha\%)$ represents training under maximum $\alpha\%$ load variation.}
        \label{fig:obf_uncer}
\end{figure*}

The performance under the ``wait-and-see'' uncertainty setting and the corresponding robust countermeasure is shown in Fig.~\ref{fig:obf_uncer}. As the load uncertainty increases, worst-case realization of the exogenous RD-stage load can significantly increase the PSO cost. The robust training problem $\mathcal{M}_{obf/uncer}$ is effective in reducing this worst-case cost, at the expense of only a slight increase in the nominal cost relative to $\mathcal{P}_{basic}$. Moreover, $\mathcal{M}_{abf}$ exhibits the poorest robust performance, since it is unaware of both the downstream optimization chain and the stressed deployment condition. 

A notable finding is that the worst-case cost associated with the ``True'' solar benchmark can exceed that of $\mathcal{M}_{obf}$ when load uncertainty exists. This is because the ``True'' benchmark is optimal only for the nominal setting in which the DP and RD stages are evaluated under matched information. Once exogenous load uncertainty is introduced at the RD stage, the relevant objective is no longer the nominal end-to-end cost, but the worst-case recourse cost under the stressed deployment condition. In this setting, a perfectly accurate renewable signal may induce a first-stage dispatch that is economical nominally but fragile to adverse load realizations, whereas the forecaster learned by $\mathcal{M}_{obf}$ may induce a more robust dispatch that remains more economical under the altered downstream operating regime. The robust counterpart $\mathcal{M}_{obf/uncer}$ further improves worst-case performance because it is trained explicitly against this exogenous RD-stage load uncertainty. In summary, perfect knowledge of one forecasted quantity does not generally imply optimal end-to-end operational performance when other uncertainties, recourse actions, or deployment-time mismatches remain.

\section{Conclusion}\label{sec:conclusion}

This paper proposed \emph{Learning-Augmented Power System Operations} (LAPSO), a unified optimization-centered framework for integrating machine learning into power-system operations. The key idea is to treat ML not only as a modeling tool, but as an explicit component of operational optimization whose design should be guided by downstream system-level outcomes. Under this view, LAPSO unifies two important coupling modes: decision-independent predictors that parameterize downstream optimization, and decision-dependent learned surrogates that enter optimization as auxiliary constraints or penalties. Based on this unified view, the paper introduced an optimization-aware design and validation framework organized around four general metric axes: solution-quality indicators, computational tractability, constraint satisfaction, and economic performance, which complement conventional predictive metrics in standalone ML pipeline.

The framework was instantiated on two representative applications, namely stability-constrained optimization and objective-based forecasting, demonstrating different applications follow the same design pipeline but different preferences. We further illustrated a hybrid forecast--operation--control setting and used the LAPSO view to separate uncertainty propagation in predicted or exogenous quantities from transferability across operational settings. To support reproducibility and future research, we released the open-source packages \texttt{gridforge} and \texttt{lapso}, which support scalable testbed generation and modular integration of learned components into existing power-system optimization models.

\appendix

\subsection{Scalability Analysis on SCO-related Experiment}\label{app:scale_sco}

\subsubsection{Settings}

As shown in Section~\ref{sec:sim_sco}, the computational time of $\mathcal{P}_{sco}$ is strongly influenced by the number of binary variables, which grows with network size and the complexity of the NN stability assessor. In addition, solver performance depends on the tightness of the bounds used in the MIL encoding of the NN (e.g., bounds on intermediate layer outputs) and on how frequently the stability constraint becomes active under different renewable-penetration regimes. 


All simulations on IEEE 39-, 57-, 118-, and 300-bus systems follow the same settings as in the 14-bus case study in Section~\ref{sec:case_study_sco}, using the \emph{complete} network-constrained UC formulation 
over a 24-hour horizon. Thanks to the proposed \texttt{gridforge} package, new grid configurations with compatible load and renewable profiles can be easily obtained. The detailed specifications of $\mathcal{P}_{basic}$ and the structures of NN stability assessor are summarized in Table~\ref{tab:scalability_opt} and Table~\ref{tab:scalability_net}, respectively. Note that the number of NN parameters is dependent on the size of the input features and the number of binary variables in $\mathcal{P}_{sco}$ is scaled as,
\begin{equation*}
    n_{Binary} = (n_g\times 3 + n_{NN}) \times 24
\end{equation*}
where the linear coefficient of 3 represents the on/off, start-up, and shut-down status for each generator and and $n_{NN}$ denotes the number of binary variables introduced by the NN encoding per time period (e.g., the number of ReLU units in the piecewise-linear network).

In the IEEE 14-bus case study (Table~\ref{tab:sco}), $\mathcal{P}_{sco}$ is solved with \texttt{MOSEK} over one year of data. In our larger-scale experiments, however, \texttt{MOSEK} often fails to converge reliably even for the IEEE 39-bus system, due to the increased number of binary variables. Therefore, we use \texttt{GUROBI} as the solver backend for the scalability experiments and report performance averaged over five randomly selected test instances. A time limit of 3600~s is imposed for each solve.

\begin{table}[h]
    \centering
    \footnotesize
    \renewcommand{\arraystretch}{1.2} 
    \caption{Optimization specification of $\mathcal{P}_{basic}$.}
    \begin{tabular}{ccccccc}
        \toprule
        & Gen. & Line & Ren. & Var. & Binary Var. & Cons. \\
        \midrule
        \textbf{14-Bus} & 5 & 20 & 4 & 1325 & 360 & 2938 \\
        \midrule
        \textbf{39-Bus} & 10 & 46 & 7 & 2746 & 720 & 6164\\
        \midrule
        \textbf{57-Bus} & 7 & 80 & 11 & 3871 & 504 & 9374\\
        \midrule
        \textbf{118-Bus} & 54 & 186 & 23 & 12630 & 3888 & 27180\\
        \midrule
        \textbf{300-Bus} & 69 & 411 & 60 & 22581 & 4968 & 51762 \\
        \bottomrule
    \end{tabular}
    \label{tab:scalability_opt}
\end{table}

\begin{table}[h]
    \centering
    \footnotesize
    \renewcommand{\arraystretch}{1.2} 
    \caption{NN specification of $\mathcal{M}_{sco}$ for scalability analysis. $NN_{L}^{\psi}$ represents neural network with $L$ layers and $\psi$ number of trainable parameters.}
    \begin{tabular}{cccccc}
        \toprule
        & 14-Bus & 39-Bus & 57-Bus & 118-Bus & 300-Bus \\
        \midrule
        \textbf{T}iny & $NN_{3}^{221}$ & $NN_{3}^{301}$ & $NN_{3}^{311}$ & $NN_{3}^{901}$ & $NN_{3}^{1421}$ \\
        \midrule
        \textbf{S}mall & $NN_{3}^{1261}$ & $NN_{3}^{1501}$ & $NN_{3}^{1531}$ & $NN_{3}^{3301}$ & $NN_{3}^{4861}$ \\
        \midrule
        \textbf{M}ed. & $NN_{3}^{3101}$ & $NN_{3}^{3501}$ & $NN_{3}^{3551}$ & $NN_{3}^{6501}$ & $NN_{3}^{9101}$ \\
        \midrule
        \textbf{L}arge & $NN_{4}^{5651}$ & $NN_{4}^{6051}$ & $NN_{4}^{6101}$ & $NN_{4}^{9051}$ & $NN_{4}^{11651}$  \\
        \bottomrule
    \end{tabular}
    \label{tab:scalability_net}
\end{table}

\subsubsection{Optimization-aware Active Dataset Sampling}\label{app:active_sampling}

The uniform sampling strategy used to construct the SCO training dataset in Section~\ref{sec:case_study_sco} becomes computationally intractable for larger systems because the number of operating-point combinations grows exponentially with the number of commitment binaries and discretized renewable levels. For example, even for the IEEE 39-bus system, enumerating the full sampling space would require approximately $(2^{10}-1)\times 5^{7} \approx 8\times 10^7$ samples, which is inefficient and impractical. Moreover, exhaustive uniform sampling is not necessary in practice: many combinations correspond to unrealistic operating conditions or lie far from the stability boundary that is most informative for learning.

Recent studies have shown that samples concentrated near classification boundaries can be highly informative for learning, while such operating points may be rare under purely economic dispatch in practice \cite{xu2024incorporation, chu2025coordinated, jia2025converter}. This motivates an \emph{optimization-aware} data-generation procedure aligned with the LAPSO pipeline (Fig.~\ref{fig:learning_pipeline}), in which dataset construction is coupled to the underlying optimization model. We therefore propose an \emph{active sampling} strategy that combines gradient-based sensitivity analysis with heuristics. The algorithm initializes the dataset from operational solutions of $\mathcal{P}_{basic}$. Specifically, we construct $\mathcal{D}_{basic}=\{(\bm{u}_g^i,\tilde{\bm{p}}_r^i)\}_{i=1}^{8760}, ~ \tilde{\bm{p}}_r^i:=\bm{p}_r^i-\bm{p}_{rc}^i $ where $\bm{u}_g^i$ denotes the commitment status and $\tilde{\bm{p}}_r^i$ denotes the net renewable injection after curtailment. We then employ augmentation mechanisms to expand $\mathcal{D}_{basic}$ toward and across the stability boundary.

Since renewable injections $\tilde{\bm{p}}_r^i$ are continuous, their influence on the gSCR index can be quantified via sensitivity analysis. In implementation, this is computed by automatic differentiation in \texttt{PyTorch}. For sample $i$, we evaluate
\begin{equation}\label{eq:gscr_grad}
    \bm{g}^i = \nabla_{\tilde{\bm{p}}_r}\,\mathrm{gSCR}(\bm{u}_g^i,\tilde{\bm{p}}_r^i).
\end{equation}
We then apply a gradient-based search to generate a near-boundary counterpart across the true gSCR threshold. Concretely, if sample $i$ is stable (unstable), we perform gradient descent (ascent) on $\mathrm{gSCR}$ to progressively decrease (increase) the stability margin until the operating point crosses into the unstable (stable) region. Because the search is initialized from economically optimal operating points produced by $\mathcal{P}_{basic}$, the resulting samples remain close to economically plausible regimes, improving the practical relevance of the sampled boundary points. The gradient-based sampling procedure is summarized in Algorithm~\ref{alg:grad_sampling}. 

\begin{algorithm}[t]
    \SetKwInOut{Input}{Input}
    \SetKwInOut{Output}{Output}
    \small

    \Input{Operational dataset $\mathcal{D}_{old}$, Maximum update number $no$, Step size $lr$, Maximum Renewable Generation $\bm{p}_{r}^{max}$}
    \Output{Close-to-boundary dataset $\mathcal{D}_{new}$}

    $\mathcal{D}_{new} = \{\}$

    \For{$i=1,\dots,|\mathcal{D}_{old}|$}
    {  
        \tcc{Determine the search direction}
        $\gamma = \operatorname{sign}(\operatorname{gSCR}(\bm{u}_g^i,\tilde{\bm p}_t^i) - \operatorname{gSCR}_{lim})$
        
        \For{$k = 1:no$}
        {   
            $\tilde{\bm{p}}_r^{i,pre} = \tilde{\bm{p}}_r^{i}$
            
            \tcc{Obtain the gradient \eqref{eq:gscr_grad}}

            $\bm{g}^{i} = \nabla_{\tilde{\bm{p}}_r}\,\operatorname{gSCR}(\bm{u}_g^i,\tilde{\bm{p}}_r^i)$

            \tcc{Update}
            $\tilde{\bm{p}}_r^{i} \leftarrow \tilde{\bm{p}}_r^{i} - \gamma \cdot lr \cdot \bm{g}^{i}$

            \tcc{Project to the feasible region}
            
            $\tilde{\bm{p}}_r^{i} = \operatorname{Clip}(\tilde{\bm{p}}_r^{i}, \operatorname{min}= \bm{0}, \operatorname{max} = \bm{p}_r^{max})$

            \tcc{Termination}
            \If{$\gamma \cdot (\operatorname{gSCR}(\bm{u}_g^i,\tilde{\bm{p}}_r^i) - \operatorname{gSCR}_{lim}) < 0$}
            {
                $\mathcal{D}_{new} = \mathcal{D}_{new} \cup (\bm{u}_g^i, \tilde{\bm{p}}_r^i) \cup (\bm{u}_g^i, \tilde{\bm{p}}_r^{i,pre})$
                
                Break
            }
        }
        $\mathcal{D}_{new} = \mathcal{D}_{new} \cup (\bm{u}_g^i, \tilde{\bm{p}}_r^i) \cup (\bm{u}_g^i, \tilde{\bm{p}}_r^{i,pre})$
    }
    \caption{Gradient-based Sampling $\mathcal{GS}(\cdot)$}
    \label{alg:grad_sampling}
\end{algorithm}

Note that the generator commitment $\bm{u}_g^i$ is kept fixed when updating $\tilde{\bm{p}}_r^i$. Since $\bm{u}_g^i$ is discrete, we additionally introduce a heuristic procedure to explore samples across the stability boundary by modifying commitment patterns. Specifically, to convert an unstable sample into a stable one, we iteratively switch on an offline generator with the lowest fixed cost $\bm{c}_{fix}$, so that additional online capacity tends to increase gSCR while avoiding unnecessary cost increases. Conversely, to convert a stable sample into an unstable one, we iteratively switch off an online generator with the highest fixed cost, which preserves low-cost commitment patterns while reducing the stability margin. The heuristic-based $\bm{u}_g$ sampling is summarized in Algorithm~\ref{alg:heur_sampling}. After updating $\bm{u}_g$, the gradient-based sampling in Algorithm~\ref{alg:grad_sampling} can be applied again to refine $\tilde{\bm{p}}_r$ under the modified commitment.

To summarize, the augmentation of the training dataset for $\mathcal{M}_{sco}$ proceeds as follows:
\begin{enumerate}[leftmargin=*]
    \item Given the one-year load and renewable profiles, construct $\mathcal{D}_{basic}$ by solving $\mathcal{P}_{basic}$ with the UC formulation.
    \item Perturb $\bm{u}_g$ to obtain $\mathcal{D}_{u}=\mathcal{HS}(\mathcal{D}_{basic})$.
    \item Perturb $\tilde{\bm{p}}_r$ to obtain $\mathcal{D}_{up}=\mathcal{GS}(\mathcal{D}_{basic}\cup\mathcal{D}_{u})$.
    \item Form the final training dataset as $\mathcal{D}_{sco}=\mathcal{D}_{basic}\cup\mathcal{D}_{u}\cup\mathcal{D}_{up}$.
\end{enumerate}
Since both $\mathcal{HS}(\cdot)$ and $\mathcal{GS}(\cdot)$ generate two samples per input sample (one on each side of the boundary), $\mathcal{HS}(\mathcal{D}_{basic})$ doubles the dataset size and $\mathcal{GS}(\mathcal{D}_{basic}\cup\mathcal{D}_{u})$ doubles its input size. Consequently, $|\mathcal{D}_{sco}| = 9|\mathcal{D}_{basic}| = 78{,}840$ training samples in this case study.

\begin{algorithm}[t]
    \SetKwInOut{Input}{Input}
    \SetKwInOut{Output}{Output}
    \small

    \Input{Operational dataset $\mathcal{D}_{old}$, Number of generator $n_g$, Generator fixed cost $\bm{c}_{fix}$}
    \Output{Close-to-boundary dataset $\mathcal{D}_{new}$}

    $\mathcal{D}_{new} = \{\}$
    
    \For{$(\bm{u}_g,\tilde{\bm{p}}_r) \in \mathcal{D}_{old}$}
    {  
        \tcc{Determine the search direction}
        $\gamma = \operatorname{sign}(\operatorname{gSCR}(\bm{u}_g^i,\tilde{p}_t^i) - \operatorname{gSCR}_{lim})$
        
        \For{$k = 1:n_g$}
        {   
            $\bm{u}_g^{i,pre} = \bm{u}_g^i$
            
            \eIf{$\gamma$ < 0}
            {
                \tcc{From unstable to stable}
                $j = \arg\min_{j:\bm{u}_g[j] = 0}\bm{c}_g[j]$
                
                $\bm{u}_g^i[j] = 1$
            }{
                \tcc{From stable to unstable}
                $j = \arg\max_{j:\bm{u}_g[j] = 1}\bm{c}_g[j]$
                
                $\bm{u}_g^i[j] = 0$
            }

            \tcc{Termination Condition}
            \If{$\gamma \cdot (\operatorname{gSCR}(\bm{u}_g^i,\tilde{\bm{p}}_t^i) - \operatorname{gSCR}_{lim}) < 0$}
            {
                $\mathcal{D}_{new} = \mathcal{D}_{new} \cup (\bm{u}_g^i, \tilde{\bm{p}}_r^i) \cup (\bm{u}_g^{i,pre}, \tilde{\bm{p}}_r^i)$
                
                Break
            }
        }
        $\mathcal{D}_{new} = \mathcal{D}_{new} \cup (\bm{u}_g^i, \tilde{\bm{p}}_r^i) \cup (\bm{u}_g^{i,pre}, \tilde{\bm{p}}_r^{i})$
    }
    \caption{Heuristic-based Sampling $\mathcal{HS}(\cdot)$}
    \label{alg:heur_sampling}
\end{algorithm}

\subsubsection{Simulation Results}

\begin{figure}
    \centering
    \includegraphics[width=0.60\linewidth, clip, trim=2.2cm 2.2cm 0.cm 3.5cm]{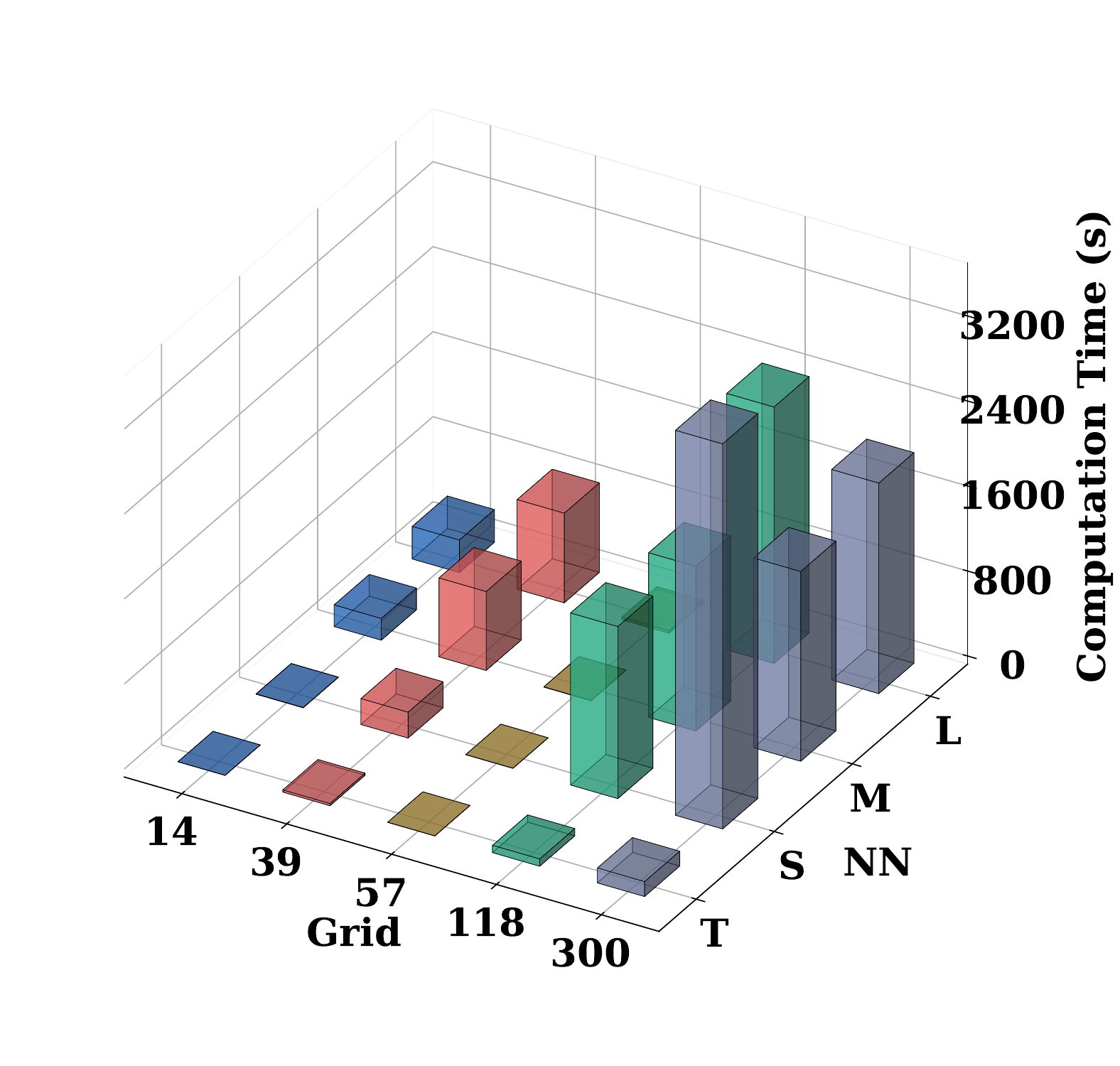}
    \caption{Computational time of $\mathcal{P}_{sco}$ against different test systems and NN structures.}
    \label{fig:computational_time_3d}
\end{figure}

Due to the computational burden, averaged performances are reported over five randomly selected load--renewable profiles. The computational times for different grid sizes and NN structures are illustrated in Fig.~\ref{fig:computational_time_3d}, and detailed results are reported in Table~\ref{tab:sco_large}.

Notably, even with more than 8{,}000 binary variables (e.g., the 300-bus system with $NN_{4}^{11656}$), $\mathcal{P}_{sco}$ can be solved within the imposed time limit in our experiments, indicating that the proposed implementation scales to the tested network sizes. This efficiency is supported by interval bound propagation (IBP), which tightens bounds on NN layer outputs and strengthens the MIL formulation. Furthermore, the stability constraints trained by the proposed active sampling strategy achieve strong stabilization performance, with SR exceeding 92\% across all instances and reaching 100\% in most settings.

\begin{figure}
    \centering
     \begin{subfigure}[b]{0.48\linewidth}
         \centering
         \includegraphics[width=\textwidth]{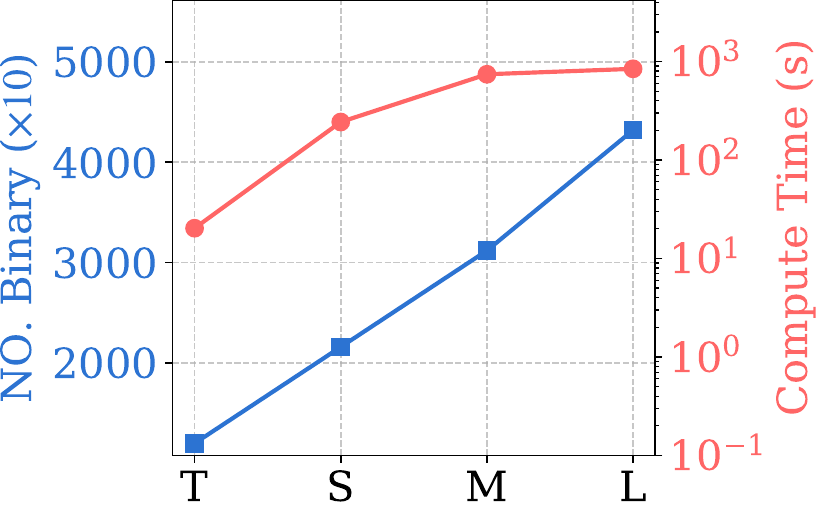}
         \caption{39-Bus}
     \end{subfigure}
     \hfill
     \begin{subfigure}[b]{0.48\linewidth}
         \centering
         \includegraphics[width=\textwidth]{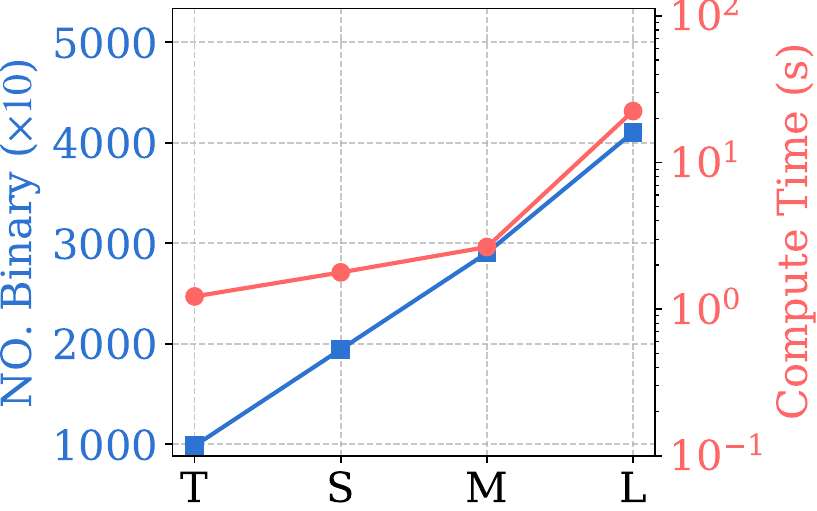}
         \caption{57-Bus}
     \end{subfigure}
     \hfill
     \begin{subfigure}[b]{0.48\linewidth}
         \centering
         \includegraphics[width=\textwidth]{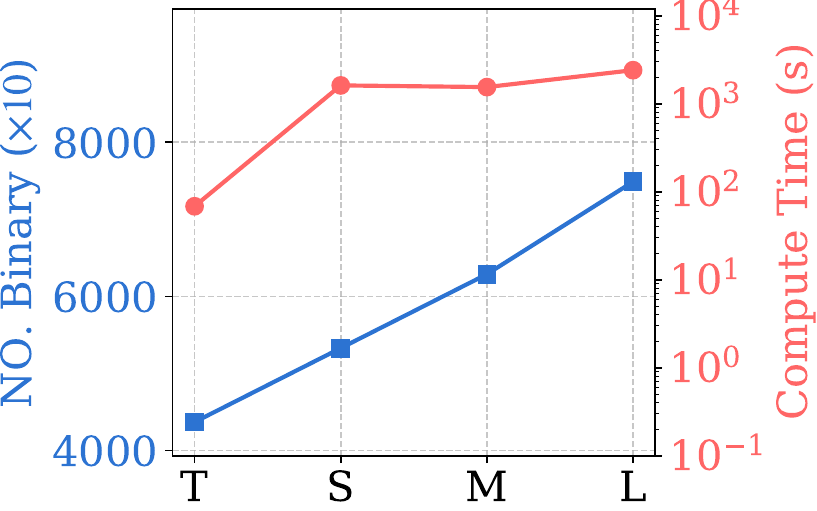}
         \caption{118-Bus}
     \end{subfigure}
     \hfill
     \begin{subfigure}[b]{0.48\linewidth}
         \centering
         \includegraphics[width=\textwidth]{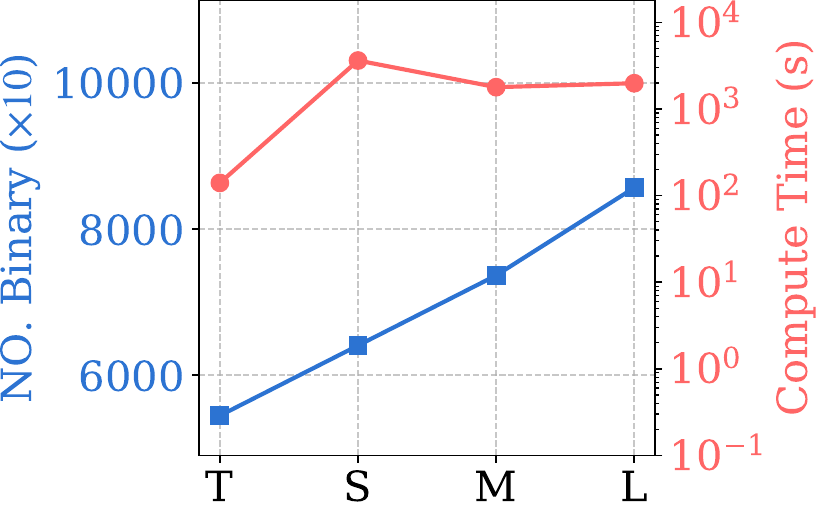}
         \caption{300-Bus}
     \end{subfigure}
        \caption{Scalability of computational performances over different number of binary variables of $\mathcal{P}_{sco}$ on different systems.}
        \label{fig:sco_compare_large}
\end{figure}

The computational time exhibits a similar scaling trend with respect to the number of binary variables as previously observed (Fig.~\ref{fig:sco_compare} and Fig.~\ref{fig:sco_compare_large}). A few outliers exist; for example, the 300-bus system with the small NN $NN_{3}^{4861}$ incurs the highest computational cost. This may be due to instance-specific solver behavior and tighter coupling between the encoded NN-based stability constraint and the remaining UC constraints. In contrast, the 57-bus system exhibits relatively low computational time, consistent with its smaller number of generators (Table~\ref{tab:scalability_opt}).

Beyond the tested 300-bus system, the main limitation of the present SCO workflow is the scalability of the underlying MILP rather than the predictive model alone. In very large systems, even the baseline full UC problem $\mathcal{P}_{basic}$ is already computationally demanding; once encoded NN-based stability constraints and repeated solves for active sampling are added, both deployment-time optimization and training-time data generation can become prohibitive.

\begin{table*}[h]
    \centering
    \footnotesize
    \caption{Scalability Performance of SCO with different stability assessors, averaged over 5 randomly picked days. \texttt{GUROBI} is used as the solver backend.}
    \begin{tabular}{c|c|c|c|c|c|c|c|c|c|c|c|c|c}
         \multicolumn{2}{c|}{} & \multicolumn{3}{c|}{\textbf{14-Bus System}} & \multicolumn{3}{c|}{\textbf{39-Bus System}} & \multicolumn{3}{c|}{\textbf{118-Bus System}} & \multicolumn{3}{c}{\textbf{300-Bus System}}  \\\hline
         \multicolumn{2}{c|}{ \textbf{Type}} & \textbf{SR}(\%) &\textbf{No.Bin.} & \textbf{Time}($s$)  &  \textbf{SR}(\%) &\textbf{No.Bin.} & \textbf{Time}($s$) &  \textbf{SR}(\%) &\textbf{No.Bin.} & \textbf{Time}($s$) &  \textbf{SR}(\%) &\textbf{No.Bin.} & \textbf{Time}($s$)   \\\hline
         \multicolumn{2}{c|}{ $\mathcal{P}_{basic}$ } & NA & 360 & 0.18 & NA & 760 & 0.354 & NA & 3888 & 8.47 & NA & 4968 & 19.13 \\\hline
       \multirow{4}{*}{$\mathcal{P}_{sco}$}
        & \textbf{T} & 100.00 & 840 & 1.03 & 100.00 & 1200 & 120.29 & 92.11 & 4368 & 68.47 & 91.23 & 5448 & 140.05 \\\cline{2-14}
        &  \textbf{S} & 100.00 & 1800 & 2.93 & 100.00 & 2160 & 243.16 & 100.00 & 5328 & 1619.08 & 87.72 & 6408 & 3623.09 \\\cline{2-14}
        & \textbf{M} & 100.00 & 2710 & 204.15 & 100.00 & 3120 & 742.49 & 100.00 & 6288 & 1549.51 & 91.23 & 7368 & 1785.7 \\\cline{2-14}
        & \textbf{L} & 100.00 & 3960 & 306.08 & 100.00 & 4320 & 843.24 & 100.00 & 7488 & 2412.2 & 91.23 & 8568 & 1981.59
    \end{tabular}
    \label{tab:sco_large}
\end{table*}

\subsection{Scalability Analysis on OBF-Related Experiment}\label{app:scale_obf}

We follow the settings in Section~\ref{sec:obf_case_study} for the scalability analysis of OBF-related tasks. Due to the computational burden, results for the IEEE 39- and 57-bus systems are averaged over 5 randomly selected weeks, while the IEEE 14-bus results are averaged over 52 weeks as before. Meanwhile, $\mathcal{M}_{obf/uncer}$ and $\mathcal{P}_{obf/uncer}$ are trained and evaluated with a 5\% load-uncertainty budget. Moreover, since $\mathcal{M}_{obf}$ for the IEEE 118-bus system does not converge within 1 hour, it is omitted from the summary.

\begin{table}[h]
    \centering
    \footnotesize
    \caption{Scalability Analysis on OBF-related tasks.}
    \begin{tabular}{c|c|c}
        \textbf{System} & \textbf{Training Method} & \textbf{Training Time (s)} \\\hline
        \multirow{4}{*}{\textbf{14-Bus}} & True & NA \\
        & $\mathcal{M}_{abf}$ & <1s \\
        & $\mathcal{M}_{obf}$ & 4.27s \\
        & $\mathcal{M}_{obf/uncer}$ & 24.83s  \\\hline
        \multirow{4}{*}{\textbf{39-Bus}} & True & NA \\
        & $\mathcal{M}_{abf}$ & <1s \\
        & $\mathcal{M}_{obf}$ & 97.17 \\
        & $\mathcal{M}_{obf/uncer}$ & 267.99 \\\hline
        \multirow{4}{*}{\textbf{57-Bus}}  & True & NA\\
        & $\mathcal{M}_{abf}$ & <1s \\
        & $\mathcal{M}_{obf}$ & 179.40 \\
        & $\mathcal{M}_{obf/uncer}$ & 372.83 \\
    \end{tabular}
    \label{tab:obf_time_scalability}
\end{table}

\begin{figure}
     \centering
     \begin{subfigure}[b]{0.49\linewidth}
         \centering
         \includegraphics[width=\textwidth]{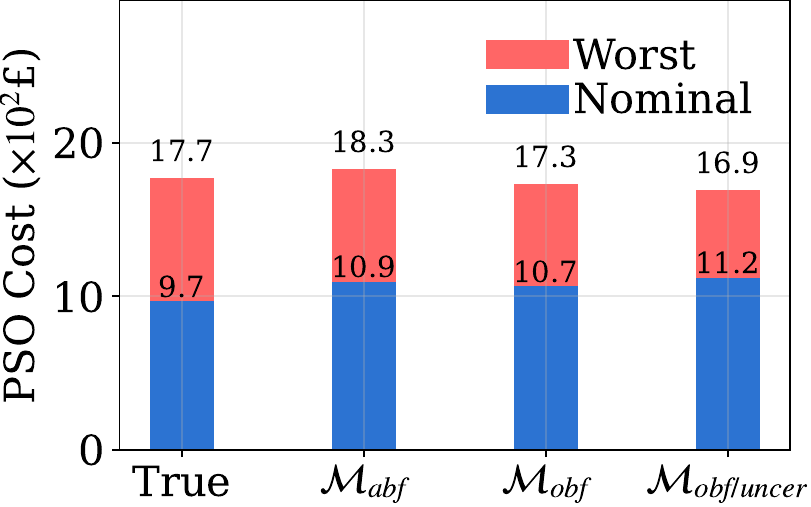}
         \caption{39-Bus}
     \end{subfigure}
     \hfill
     \begin{subfigure}[b]{0.49\linewidth}
         \centering
         \includegraphics[width=\textwidth]{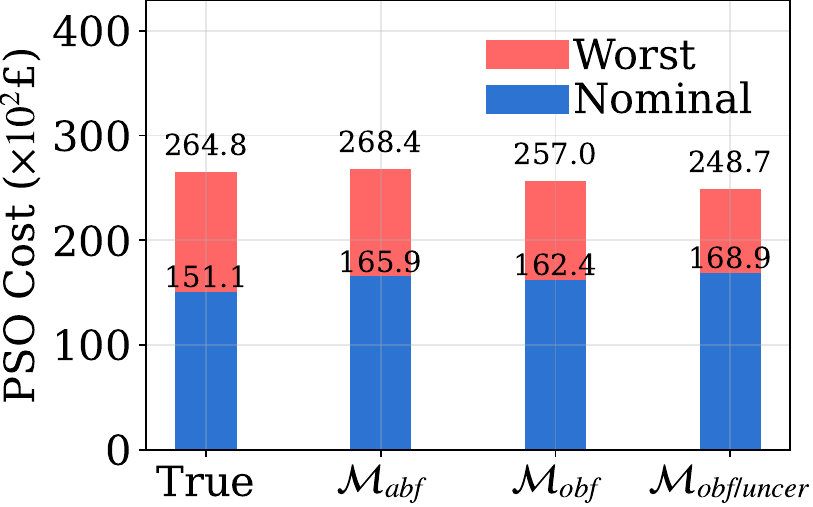}
         \caption{57-Bus}
     \end{subfigure}
        \caption{Scalability performances of different OBF training algorithms under 5\% load variation level. The $\mathcal{M}_{obf/uncer}$ is under the same 5\% uncertainty level.}
        \label{fig:obf_uncer_large}
\end{figure}

Similar to the SCO case, the computational time for $\mathcal{M}_{obf}$ and $\mathcal{M}_{obf/uncer}$ is strongly affected by the number of binary variables introduced by the KKT-based reformulation. In particular, big-$M$ linearization of complementarity conditions introduces additional binaries, and the total number of binaries grows with the number of constraints and the training horizon (both in $\mathcal{M}_{obf}$ and in the main/subproblems of $\mathcal{M}_{obf/uncer}$). As a weekly horizon of 168 hourly samples is considered in the fine-tuning stage, this growth can be significant. Nevertheless, the average training times reported in Table~\ref{tab:obf_time_scalability} remain practical for the system sizes considered in this appendix. Finally, the operational costs for the IEEE 39- and 57-bus systems are illustrated in Fig.~\ref{fig:obf_uncer_large}, demonstrating improved robustness against worst-case load variations with $\mathcal{M}_{obf/uncer}$.

\subsection{Uncertainty Analysis on Both ML and Opt Uncertainties}
\label{app:multi_uncertainty}

To further illustrate end-to-end uncertainty tracing under the unified LAPSO view (Section~\ref{sec:obf-uncertain}), this appendix considers a worst-case stress test in which \emph{both} the forecaster input (ML-Uncertainty) and the exogenous optimization input at the RD stage (Opt-Uncertainty) are uncertain simultaneously. Developing a robust training countermeasure that jointly hedges multiple uncertainty sources is left for future work.

Based on \eqref{eq:obf-case-study} and the RD worst-case formulation, under both ML and Opt Uncertainties, the worst-case end-to-end operational cost for a given sample $i$ can be written as the following robust bilevel problem:
\begin{equation}\label{eq:obf-case-worst_multi_source}
    \begin{aligned}
         \max_{\bm{y}_{1,rd}^i\in\mathcal{Y}^i_1,\; \bm{x}^i \in \mathcal{X}^i}\;\;
         & \tilde{f}_{dp}(\hat{\bm{p}}_g^i) + f_{rd}(\Delta \hat{\bm{p}}_g^i, \hat{\bm{z}}_{rd}^i) \\
        \text{s.t.}\;\;
        & (\Delta \hat{\bm{p}}^i_g, \hat{\bm{z}}^i_{rd}) \in \mathcal{KKT}_{rd}(\bm{y}^i_2, \bm{y}^i_{1,rd}, \hat{\bm{p}}_g^{i}) \\
        & (\hat{\bm p}_g^i, \hat{\bm z}_{dp}^i) \in \mathcal{KKT}_{dp}(\hat{\bm{y}}^i_2, \bm{y}_1^i) \\
        & \hat{\bm y}_2^i \in \mathcal{Y}_2^i,~~ \hat{\bm{y}}_2^i = \bm{W}_h^\star \bm{x}^i + \bm{b}_h^\star,
    \end{aligned}
\end{equation}
where $\mathcal{X}^i$ denotes the uncertainty set for the forecaster input feature (the latent embedding by Fig.~\ref{fig:obf_exp_framework}); $\mathcal{Y}_1^i$ and $\mathcal{Y}_2^i$ denote the uncertainty sets for the RD-stage exogenous load and the predicted renewable generation, respectively. $\mathcal{KKT}_{dp}(\cdot)$ and $\mathcal{KKT}_{rd}(\cdot)$ represent the KKT systems of the DP and RD subproblems, respectively, and $(\bm{W}_h^\star,\bm{b}_h^\star)$ are fixed trained forecaster parameters. In implementation, \texttt{lapso.optimization.form\_kkt} is used to generate the KKT constraints and associated variables/parameters needed to construct \eqref{eq:obf-case-worst_multi_source}. 

\begin{figure}
     \centering
     \begin{subfigure}[b]{0.49\linewidth}
         \centering
         \includegraphics[width=\textwidth]{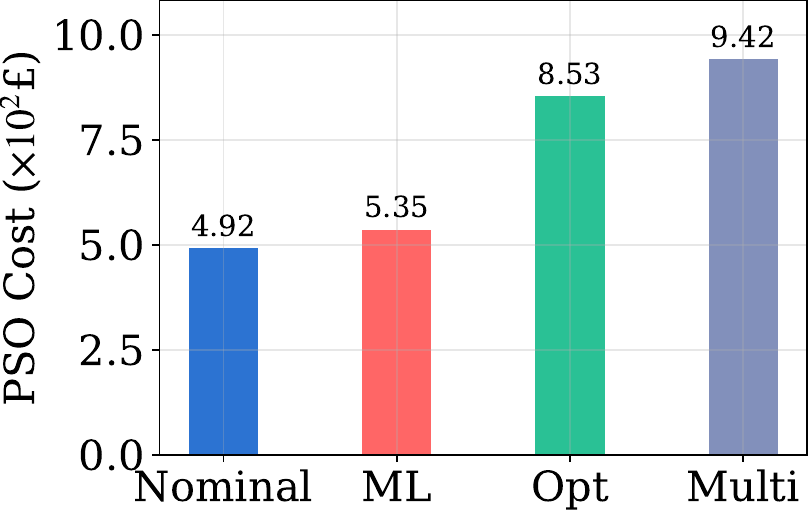}
         \caption{$\mathcal{M}_{abf}$}
     \end{subfigure}
     \hfill
     \begin{subfigure}[b]{0.49\linewidth}
         \centering
         \includegraphics[width=\textwidth]{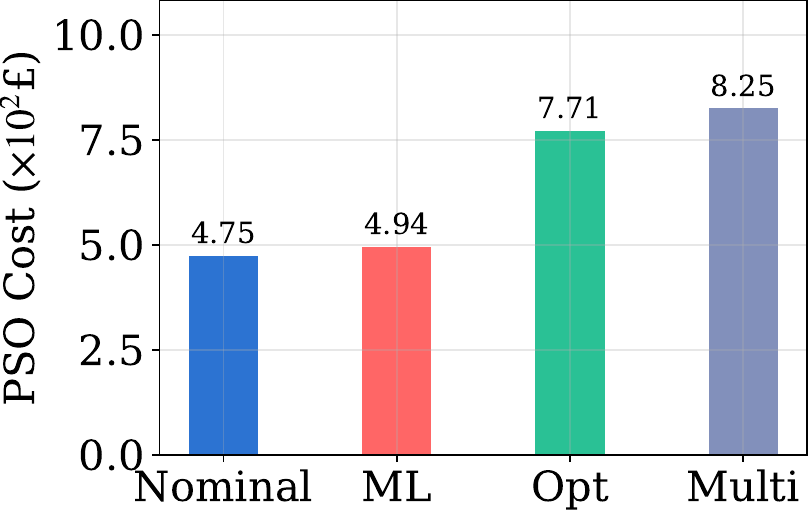}
         \caption{$\mathcal{M}_{obf}$}
     \end{subfigure}
        \caption{Performances of forecasters with Original Cost, ML-Uncertainty, Opt-Uncertainty, and Multi-Uncertainty.}
        \label{fig:obf_uncer_multi}
\end{figure}

Fig.~\ref{fig:obf_uncer_multi} reports the worst-case PSO costs for forecasters $(\bm{W}_h^\star,\bm{b}_h^\star)$ trained by $\mathcal{M}_{abf}$ and $\mathcal{M}_{obf}$ under four conditions: nominal, ML-Uncertainty only, Opt-Uncertainty only, and combined (multi-source) uncertainty. Costs are averaged over the 52 weekly test windows. In this experiment, the uncertainty sets $\mathcal{X}^i$, $\mathcal{Y}_1^i$, and $\mathcal{Y}_2^i$ are constructed using elementwise $\pm10\%$, $\pm5\%$, and $\pm5\%$ perturbations around the corresponding nominal values of sample $i$, such as the nominal NN embedding, nominal load profile, and nominal renewable forecast produced by the ABF model, respectively. Relative to the Opt-Uncertainty-only setting, adding ML-Uncertainty further increases the averaged worst-case cost by 10.4\% for $\mathcal{M}_{abf}$ and 7.0\% for $\mathcal{M}_{obf}$, indicating that compounding uncertainty sources can materially worsen worst-case system outcomes compared to single-source uncertainty.

\IEEEpeerreviewmaketitle

\bibliographystyle{IEEEtran}
\bibliography{IEEEabrv,Reference.bib}

\bstctlcite{IEEEexample:BSTcontrol}


\ifCLASSOPTIONcaptionsoff
  \newpage
\fi
\end{document}